\documentstyle[prl,aps,floats,epsfig,amssymb]{revtex} 

\begin{document}

\title{Study of Radiation Damage in Lead Tungstate Crystals 
Using Intense High Energy Beams}

\author{
V.A.~Batarin$^{1}$,
T.~Brennan$^{2}$,
J.~Butler$^{2}$,
H.~Cheung$^{2}$,
V.S.~Datsko$^{1}$,
A.M.~Davidenko$^{1}$,
A.A.~Derevschikov$^{1}$,
R.I.~Dzhelyadin$^{1}$,
Y.V.~Fomin$^{1}$,
V.~Frolov$^{3}$,
Y.M.~Goncharenko$^{1}$,
V.N.~Grishin$^{1}$,
V.A.~Kachanov$^{1}$,
V.Y.~Khodyrev$^{1}$,
K.~Khroustalev$^{4}$,
A.K.~Konoplyannikov$^{1}$,
A.S.~Konstantinov$^{1}$,
V.I.~Kravtsov$^{1}$,
Y.~Kubota$^{3}$,
V.M.~Leontiev$^{1}$,
V.S.~Lukanin$^{1}$,
V.A.~Maisheev$^{1}$,
Y.A.~Matulenko$^{1}$,
Y.M.~Melnick$^{1}$,
A.P.~Meschanin$^{1}$,
N.E.~Mikhalin$^{1}$,
N.G.~Minaev$^{1}$,
V.V.~Mochalov$^{1}$,
D.A.~Morozov$^{1}$,
R.~Mountain$^{4}$,
L.V.~Nogach$^{1}$,
V.A.~Pikalov$^{1}$,
A.V.~Ryazantsev$^{1}$,
P.A.~Semenov$^{1}$,
K.E.~Shestermanov$^{1}$,
L.F.~Soloviev$^{1}$,
V.L.~Solovyanov$^{1}$\footnote{deceased},
S.~Stone$^{4}$,
M.N.~Ukhanov$^{1}$,
A.V.~Uzunian$^{1}$,
A.N.~Vasiliev$^{1}$,
A.E.~Yakutin$^{1}$,
J.~Yarba$^{2}$            
\vskip 0.50cm                                                                 
\centerline{BTeV electromagnetic calorimeter group}   
\vskip 0.50cm                                                                 
}

\date{\today}
\address{
$^1$Institute for High Energy Physics, Protvino, Russia\\
$^2$Fermilab, Batavia, IL 60510, U.S.A.\\
$^3$University of Minnesota, Minneapolis, MN 55455, U.S.A.\\
$^4$Syracuse University, Syracuse, NY 13244-1130, U.S.A.\\
}

\maketitle
\vskip 0.5cm

\begin{abstract}{
We report on the effects of radiation on the light output of lead tungstate
crystals. The crystals were irradiated by pure, intense high energy
electron and hadron beams
as well as by a mixture of  hadrons, neutrons and gammas. The crystals were 
manufactured in Bogoroditsk,
Apatity (both Russia), and Shanghai (China). These studies were carried out
at the 70-GeV proton
accelerator in Protvino.}
\end{abstract}
\vskip 0.5cm

\section{Introduction}
    The BTeV~\cite{proposal} experiment is being readied to study beauty and charm physics 
at the Fermilab Tevatron collider. The goals are to make an exhaustive search 
for physics beyond the Standard Model (SM) and make precise 
measurements of the SM parameters. The important measurements to make 
involve CP 
violation, mixing, and rare decays of hadrons containing b or c quarks. Since
detection of photons, mostly from $\pi^o$ or $\eta$ decays is essential
to accomplish our physics objectives, we have decided to use an 
electromagneticcalorimeter (EMCAL) made
of lead tungstate PbWO$_4$ (PWO) crystals. These crystals produce light proportional 
to the incident electromagnetic energy; this light will be sensed by photomultiplier tubes. 
This system is ideal for a heavy quark experiment at a hadron collider 
because of excellent energy and position resolution, a compact shower size that minimizes 
overlapping showers (due to the  small Moliere radius), fast signals that minimize 
shower overlaps in time and expected excellent radiation hardness.

Pioneering work on PWO crystals performance was done at Protvino~\cite{protvino}. 
These results showed the promise of such crystals. However, the technology of mass producing 
such crystals with high purity was not yet known. The CMS group worked with companies 
both in Russia and China to perfect these techniques~\cite{TDR}.

In high luminosity collider experiments, PWO crystals will be irradiated
by high energy particles and accumulate significant absorbed doses, up to a few Mrad. 
The radiation hardness of PWO crystals has been studied by the CMS group using radioactive sources and electron beams~\cite{cmsn}. The general 
conclusion is that lead tungstate crystals were radiation hard, and that the damage in crystals depends only on the dose rate~\cite{cms_conclusion}.
It is, however, important to measure radiation damages of PWO crystals in high energy particle environments which are more similar to that which these 
crystals will be exposed to.

It is important to emphasize \cite{uzu,huh} that in a hadron collider 
experiment radiation effects from hadronic interactions and neutrons 
could be much more serious than 
seen with photons or electrons of the same doses.
Compared to photons or electrons, high-energy
hadrons will be able to induce inelastic nuclear reactions which
will locally destroy the crystal lattice. In particular, they can create
nuclear fragments with very high energy transfer and lead to extended
clusters of crystal lattice distortion. A simple calculation suggests that
such interaction may produce significant number of additional crystal defects
over the life of BTeV. 
Therefore it is crucial to study the radiation hardness of PWO-crystals 
using a hadron environment which is similar to the BTeV EMCAL
expectations. Such radiation studies with lead tungstate
crystals have been carried out for the first time. The results of this
study are presented in this paper.

     The general goal of our test beam studies was to evaluate 
the performance of lead tungstate crystals produced by two manufacturers in Russia, Bogoroditsk 
and Northern Crystal in Apatity, and one in China, Shanghai Institute of Ceramics. More specific goals were to understand how to set specifications 
for purchasing crystals, confirm energy and position resolution predictions,
measure the radiation rate dependence of light output, and 
measure the correlation between light output and the LED calibration system at
varying radiation loads. The 2B beam channel at the Protvino accelerator U70
has been specifically developed to provide these measurements~\cite{nim1}.
Results on energy and position resolutions of the PWO crystals which were 
obtained in these runs have been published elsewhere~\cite{nim2}.

This paper is organized as follows. A general picture of radiation damage
of PWO crystals as well as the results of simulations on dose
rate profiles in the PWO crystals with the use of the MARS program~\cite{mars} 
are described in Sec. II. These calculations are made for the BTeV 
experiment and for the two 
types of radiation studies of PWO crystals which have been carried out 
in Protvino for BTeV. In these studies we  irradiated crystals with (a)  
moderate dose rates (1-60 rad/h) of high-intensity high-energy electron
and pion beams in the secondary 
particle channel 2B and (b) super-intensive dose 
rates of mixed beam at a dedicated facility that was several meters away from the main ring of the U70. The test beam facility for approach (a), including
phototube monitoring as well as the results of the moderate
dose rates irradiation are discussed in Sec. III. Three accelerator runs, each up to a month long were devoted to these studies. The results from approach 
(b) are given in Sec. IV. The conclusions of the entire radiation studies are
presented in Sec. V.

\section{Radiation damage and absorbed dose profiles in the crystals}

Radiation hardness studies of detectors and electronics are an 
important concern in EMCAL design~\cite{TDR,cmsn}
All crystal scintillators suffer from radiation damage. 
The most common radiation damage
is due to color center formation, which results from trapping of electrons
in crystal defects such as vacancies, displacements and impurities~\cite{zhu}. 
These electrons are often in metastable states and can be excited by visible photons to higher energies. Color centers reduce light transparency of crystals,
resulting in reduced light output. Additional damage may be caused by hadrons
when they create crystal defects by displacing nuclei or changing
nuclei to different nuclei. This kind of damage can not only reduce light
transparency, but, in principle, also reduce primary scintillation light 
itself. It would be more difficult to monitor the latter effect.
Since the trapped electrons are in metastable states of
varying lifetimes and ``potential barriers'', some of them may disappear very
quickly, whereas others may be almost permanent.

When a PWO crystal no longer receives radiation, its
color centers (semi-stable excited states) disappear, and it recovers 
from transmission degradation by natural room-temperature annealing.
In fact, this annealing goes on even during radiation exposure. In general
the rate of radiation damage decreases with the amount of damage. Therefore,
when crystals are exposed to a constant radiation level, they lose 
light only up to the point when the rates of radiation damage and
natural recovery balance. Raised temperatures accelerate the recovery process and so may ultra violet irradiation. Because the damage may recover at room
temperature, it leads to a dose rate dependence of the light output.\\ 

The CMS experimental data,  mainly from photon and electron 
irradiation, indicate that the light transmission of crystals deteriorates 
due to formation of color centers by radiation, while the scintillation 
mechanism itself seems unaffected. 
Besides dependence on the dose rate, the radiation damage of PWO crystals
could also be sensitive to the type of radiation. 
In particular, the properties of crystals could be significantly 
degraded in hadron beams by displacement damage effects, i.e. distortions of the crystal structure. In these studies it is very useful to know the hadron 
fluence, the hadron spectra
and the absorbed dose rate. \\

      The BTeV EMCAL extends radially outward from the beam line. The 
crystals near the beam pipe receive the maximum dose. In order to 
ascertain the level of radiation in the 
crystals we performed calculations using the MARS code. Results are given 
in Table~\ref{tab:summary}.\\ 

\begin{table}[htb]
\caption{Fraction of BTeV crystals with given absorbed doses and dose rates 
estimated at the maximum of the dose profiles inside the crystals 
(100 rad = 1 Gy) }
\label{tab:summary}
   \begin{center}
     \begin{tabular}{|c|p{4cm}|p{4cm}|} \hline
Fraction    & Absorbed dose & Dose rate  \\
  (\%)      & (krad/year)  &  (rad/h)    \\ \hline \hline
11         & 0.3 - 1     &  0.11 - 0.36    \\ \hline
22         & 1 - 2     &  0.36 - 0.72    \\ \hline
27         & 2 - 5     &  0.72 - 1.8    \\ \hline
12         & 5 - 10     &  1.8 - 3.6    \\ \hline
16         & 10 - 50    &  3.6 - 18    \\ \hline
6         & 50 - 100     &  18 - 36    \\ \hline
3         & 100 - 200     &  36 - 72    \\ \hline
2         & 200 - 500     &  72 - 180    \\ \hline
0.4         & 500 - 1000     &  180 - 360    \\ \hline
0.2         & 1000 - 2000     &  360 - 720    \\ \hline
      \end{tabular}
    \end{center}
\end{table}

\begin{figure}[p]
\centerline{\psfig{figure=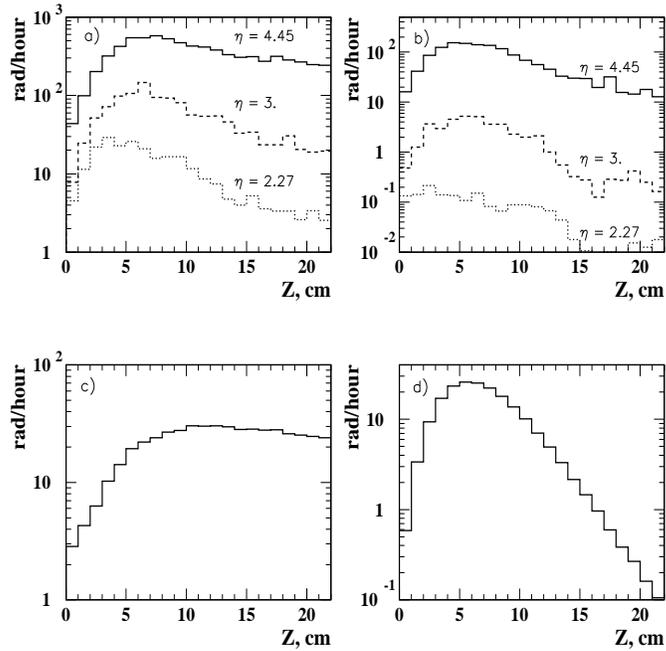,width=100mm,height=100mm}}
\caption{ Longitudinal profiles of the absorbed dose rate at the vertical (a) and 
horisontal (b) planes of the BTeV EMCAL at different rapidities, and at IHEP 
testbeam with 40 GeV pions (c) and 27 GeV electrons (d). The length of the
crystal is 22 cm. The electron profile is normalized by 10$^4$ e$^-$/sec,
and the pion profile by 10$^5 \pi$/sec.} 
\label{fig:btev_testbeam}
\end{figure}
\begin{figure}[p]
\centerline{\psfig{figure= 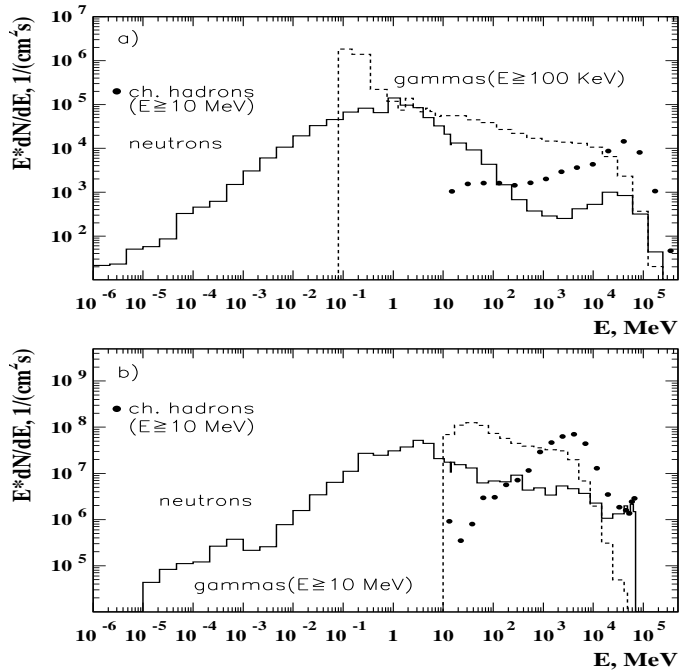,width=100mm,height=100mm}}
\caption{(a) Particle spectra at the BTeV EMCAL. (b) Particle spectra in the 
dedicated superintensive dose zone near the vacuum ring of the U-70 
accelerator. These spectrum shapes are very similar although (b) is about
three orders of magnitude higher.}
\label{fig:particle_spectra}
\end{figure}
\begin{figure}[ht]
\centerline{\psfig{figure= 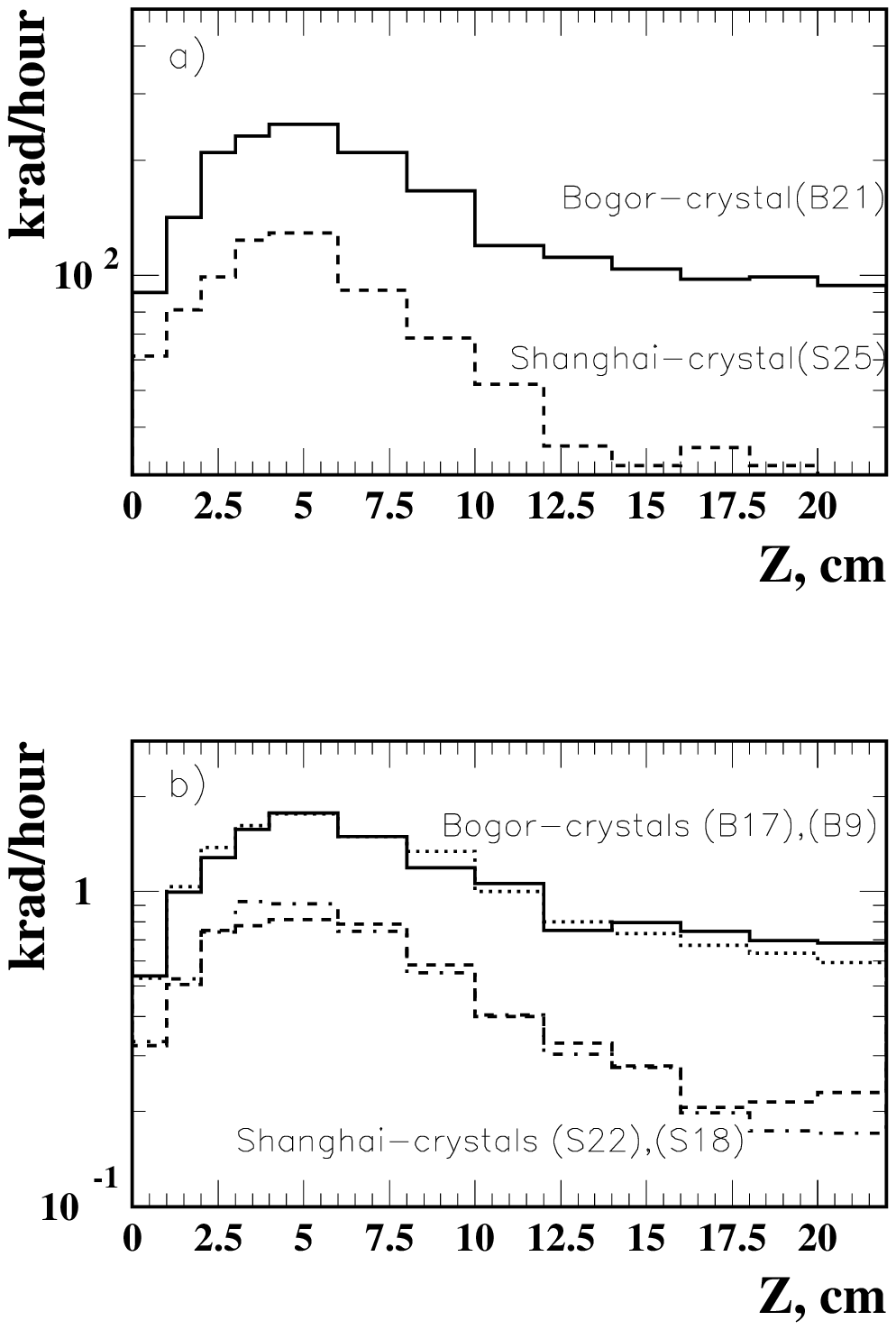,width=100mm,height=100mm}}
\caption{ Absorbed dose rates as a function of longitudinal position at the dedicated
facility near the internal target 27 of the U-70 accelerator for two crystals 
in the exposure (a) and four other crystals in the exposure (b). The intensity 
of primary 70-GeV protons at the internal target in the second exposure was 
three orders of magnitude less than in the first one.} 
\label{fig:crazy_profiles}
\end{figure}

We tried to emulate the BTeV conditions as much as possible. 
A 27 GeV electron beam and a 40 GeV $\pi^-$ beam have been used 
to irradiate the
crystals with moderate dose rates. The beams were directed into the secondary
beam channel from the accelerator, where 
primary 70-GeV protons interacted with an internal target. The MARS
calculations of the absorbed dose rates in
the crystals from the secondary beam channel are compared with the absorbed
dose rates expected in BTeV in Fig.~\ref{fig:btev_testbeam}. The $\eta$
(pseudo-rapidity) shown here reflects the coverage of the BTeV EMCAL, where
$\eta$ of 4.45 is at the extreme inside near the beam and $\eta$ of 2.27 is on
the extreme outside. Electron and pion dose profiles in the crystals are
different. The crystals receive damage from pions almost uniformly along
their length starting from a distance of 5-7 cm from the front. For
electrons an absorbed dose rate at shower maximum is two orders of
magnitude higher than near the crystal ends. Because the BTeV dipole magnet 
sweeps particles vertically, the radiation profile at the calorimeter is 
different in the horizontal and vertical planes.  Thus in BTeV the mix of 
charged hadrons and photons changes and the ratio between shower maximum and the crystal ends is only a few times in the vertical plane and an order
of magnitude in the horizontal plane. That is why both electron and pion
beams are used to study radiation damage of the crystals.\\

   Two crystals, one manufactured in Bogoroditsk and the other in Shanghai were
placed near the vacuum pipe of the Protvino U-70 accelerator in the
first dedicated super-intensive dose rate study. These crystals were irradiated by 
secondary particles coming out the internal target of the accelerator. 
The energy spectra of neutrons, gamma-quanta and charged hadrons at the place where 
the crystals were irradiated are shown in 
Fig.~\ref{fig:particle_spectra}(b). For comparison the expected particle 
spectra at the front face of the BTeV EMCAL are presented at the top part 
of the same Figure. We can see that the spectra look similar, although the 
dose rate in the IHEP irradiation zone is about three orders of magnitude 
higher than expected in BTeV. In the second dedicated intensive study, 
four more crystals from Bogoroditsk and Shanghai were exposed to radiation at the same 
facility. The intensity of the second run was reduced by two orders 
of magnitude. Absorbed dose rates as a function of longitudinal position for these two exposures are presented in Fig.\ref{fig:crazy_profiles}.\\

\section{Moderate dose rate irradiation}

In this Section, we describe the testbeam facility for moderate dose rate irradiation 
studies, discuss 
phototube gain monitoring, and present the results of irradiating 
crystals with electrons and pions.
In our radiation studies we wanted to use radiation conditions as close to the 
BTeV conditions for the crystals as possible. Absorbed dose rates as a 
function of longitudinal profile at the BTeV EMCAL 
and at IHEP testbeam have been already discussed
in Section II and presented
in Fig.~\ref{fig:btev_testbeam}. We used 27-GeV electrons and 40-GeV
pions to irradiate crystals in the three accelerator runs.

\subsection{Test beam facility}

The test beam setup 
consisted of 5x5 PWO crystal array
situated inside a temperature controlled light-tight box (ECAL), 
a beam with a momentum tagging system and a 
scintillation counter trigger system~\cite{nim1},~\cite{nim2}.

All the crystals we used were rectangular in shape.
The Bogoroditsk and Shanghai crystals were $27\times 27$ mm$^2$ 
in cross section and 220 mm in length. The Apatity crystals
were $22\times 22$ mm$^2$ in cross section and 180 mm in length. 
Light from each crystal was collected by a 10-stage
1-inch diameter Hamamatsu R5800 photomultiplier tube (PMT).
All the crystals were wrapped
by a 170 $\mu$m thick tyvek. A radioactive source study at University of
Minnesota showed that tyvek is radiation hard up to at least a few Mrad. 
This study as well as the Belarussian State University(Minsk) one also showed that a borosilicate glass did not lose 
any light at least up to 10 krad, a quartz glass up to 1 Mrad, both with an 
accuracy of 1$\%$. Six quartz PMT's were used for a part of our test 
beam study, the rest were the borosilicate PMT's.  

We accumulated absorbed doses in our crystals up to a few krad. No changes
inside the box or PMT HV values were made during the irradiation period. 
The PWO light yield strongly depends on crystal
temperature~\cite{TDR}. The 25 crystals were surrounded by a set of four
copper plates that were water cooled, which enabled a temperature control 
using a Lauda cryothermostat. The temperature for the study described in 
this paper was fixed at $20^0C$ $\pm$ $0.1^0C$. To measure the temperature 
of the crystals,
24 temperature sensors were mounted on the front and rear faces
of the crystals.  

For the most of the results presented in this paper, the crystal array was 
monitored with the four different wave length light emitting diodes (LED).   
The LEDs had the following wavelengths : 660 nm (red), 580 nm (yellow),
530 nm (green), and 470 nm (blue). Transmission of red light in
the crystals is not affected much by radiation damage~\cite{red_led}, so the
red LED monitors the PMT gain change. 
One LED generator
with a multiplexer was placed into the light-tight box with the
crystals and used for the all the LEDs. The LED temperature
dependence is on average 1$\%$/$1^{\circ}$C, and thus limited to 0.1\% because of our careful 
temperature control inside the box.  We had one bunch of fibers between the LED generator
and the crystals. In each accelerator cycle 10 pulses
data for one LED color were collected. Four cycles were needed to collect 
all the LED signals.

\begin{figure}[ht]
\centerline{\psfig{figure= 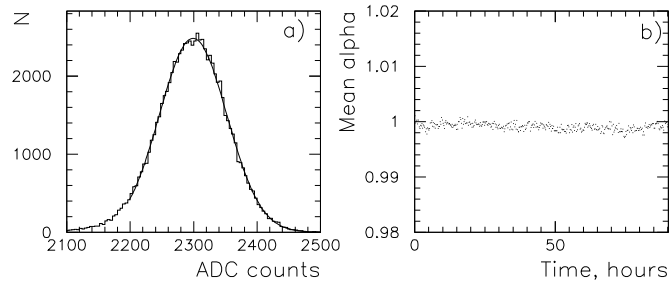,width=100mm}}
\caption{ (a) $\alpha$-spectrum accumulated over 1.5 hours. 
Sigma/mean = 2.3$\%$ when it is 
fitted by a Gaussian. (b) Normalized $\alpha$-signal in time to show  
$\alpha$-stability
 over 85 hours. Each point corresponds to
a 15-minute measurement. } 
\label{fig:alpha}
\end{figure}

      An $\alpha$-source (YAP-light pulser~\cite{alpha})   
was mounted on the photocathode
of a separate PMT in addition to the fiber to monitor LEDs themselves.  
It had 20 decays/sec
with about 5,000 photons/pulse. 
Forty pulses were collected each spill. A signal from last dynode of this PMT was
used to form an $\alpha$-trigger. The size of the YAP crystal was 3$\times$3
$mm^2$ with thickness of 0.1 mm.  Its emission spectrum has the maximum at 360 nm. The YAP crystal temperature dependence
of the light output was 0.4$\%$/$1^{\circ}$C. 
The $\alpha$-spectrum as well as $\alpha$-stability is
presented in Fig.~\ref{fig:alpha}. One can see that this stability over 85
hours is better than 0.2$\%$.
A Hamamatsu PIN diode S6468-05 with integrated amplifiers was also used to monitor 
the LEDs because it has
a good sensitivity in the red region as well as a gain stability. It's temperature 
dependence is much less than 0.4$\%$/$1^{\circ}$C. \\

We did not use an optical grease coupling between the crystals and the PMT's in order 
to avoid a contribution of a possible radiation damage of the grease. 
The PMT's were attached to the crystals without any optical material between them.
High voltage to the tubes was supplied by a LeCroy 1440 HV system.
Signals were sent to the control room patch-panel without any connection
to ground inside the crystal box to avoid ground loops. A LeCroy 2285 15-bit
integrating ADC was used to measure charge over 150 ns without
pedestal subtraction. The  
ADC sensitivity was 30 fC per count. At HV values around 1000 V in the 
tubes we had about 2 MeV/ADC count.\\

\subsection{Phototube gain change monitoring}

        We used high-intensity high-energy electron beam to irradiate the
crystals and at the same time monitor the light output. The beam
particles travel along the length of the crystals toward the PMT. We needed 
to take into account the possibile phototubes gain changes, for example, from 
varying in the beam intensity. Thus, we carried out two types 
of PMT gain change studies 
to separate the effect of PMT gain change from crystal 
radiation damage. 
We investigated the possible changes in PMT gains at a dedicated stand at IHEP 
after the accelerator runs. We also monitored the PMT's continuously during one
of the runs using the red LED.\\ 

\begin{figure}[ht]
\centerline{\psfig{figure= 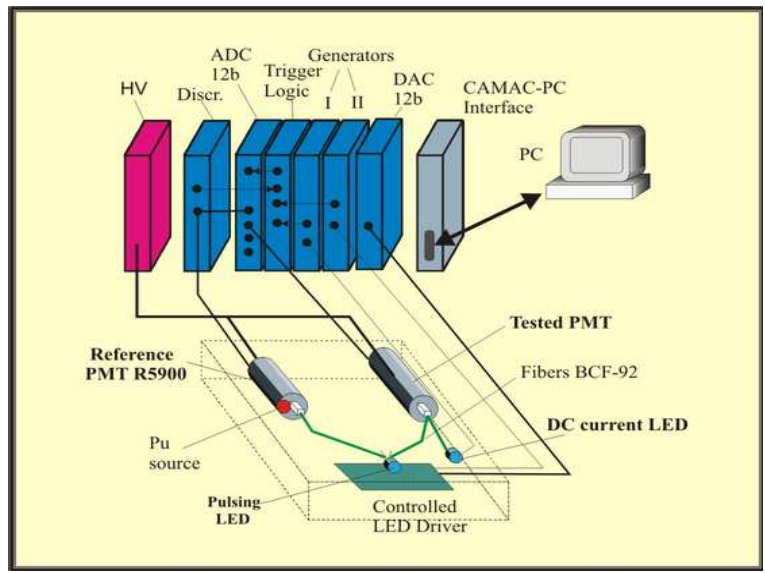,width=100mm,height=75mm}}
\caption{Sketch of a dedicated stand setup to study a PMT gain variation. }
\label{fig:lhcb_stand}
\end{figure}

\begin{figure}[bh]
\centerline{\psfig{figure= 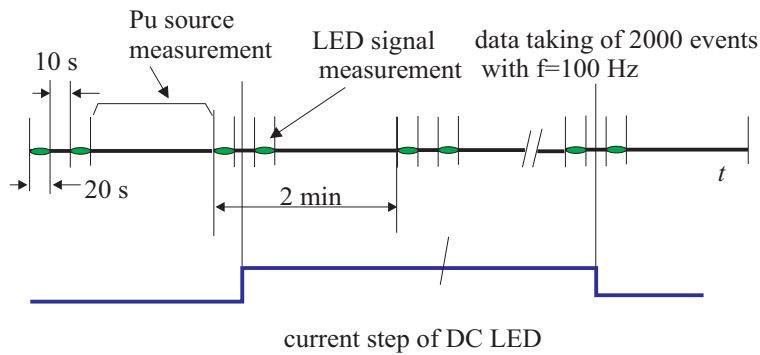,width=100mm}}
\caption{ A timing diagram of our test stand to measure PMT gain variations. }
\label{fig:lhcb_time}
\end{figure}

   Fig.~\ref{fig:lhcb_stand} shows a schematics of the dedicated stand setup to study the PMT behavior, where the average anode current was adjustable 
by changing the intensity of DC light shining on the PMT. 
The setup consisted of a high quality referenced
PMT(Hamamatsu R5900), a blue LED light pulser, a DC LED. Both pulsed and DC LED lights 
were injected into
the test PMT through optical fibers. The stability of the pulsed LED
itself was monitored by a $Pu$ radioactive source implanted in a crystal
and mounted at the photocathode of the reference PMT. The read-out
and control electronics were placed in a CAMAC crate which had
an interface with a PC. The average anode
current was chosen for each test PMT to be the
same as what we had at the test beam. The anode current was measured directly
by an ammeter. 
Fig.~\ref{fig:lhcb_time} shows a timing diagram of various measurements. Each set
of measurements took 2 minutes. At the beginning of each set, we measured the pulse
heights of two groups of 2000 light pulses. It took 20 sec to collect 2000 pulse data
and there was a 10-sec interval between the two groups.
The data from a radioactive
source in a self trigger mode were collected during the remaining 70 seconds.
This 2-minute set was then continuously repeated. The intensity of the DC
LED to induce a finite average anode current in the tested PMT
was allowed to change, if needed, in the 10 second time intervals. This system allowed us to 
make PMT long-term stability measurements with a
precision of 0.2$\%$. \\
\begin{figure}[t]
\centerline{\psfig{figure= 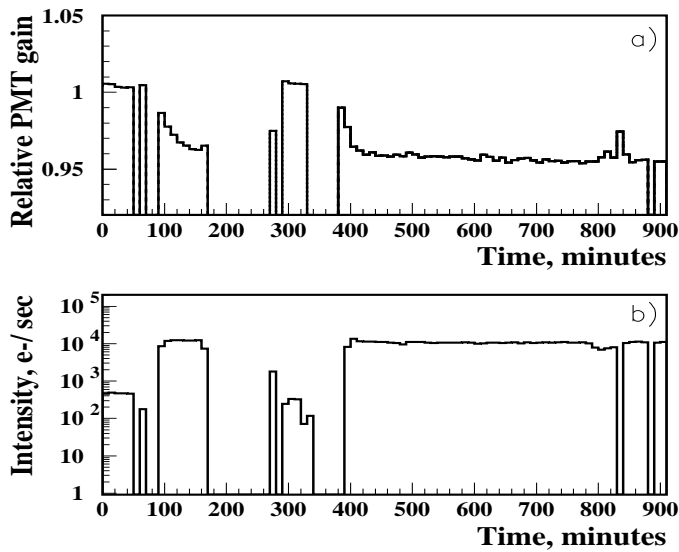,width=100mm,height=85mm}}
\caption{ (a) The response of PMT number 743 using a red LED for a 
central crystal in the array as a function of time. (b) Beam intensity 
in this counter as a function of time.}
\label{fig:kpm13}
\end{figure}

\begin{figure}[hb]
\centerline{\psfig{figure= 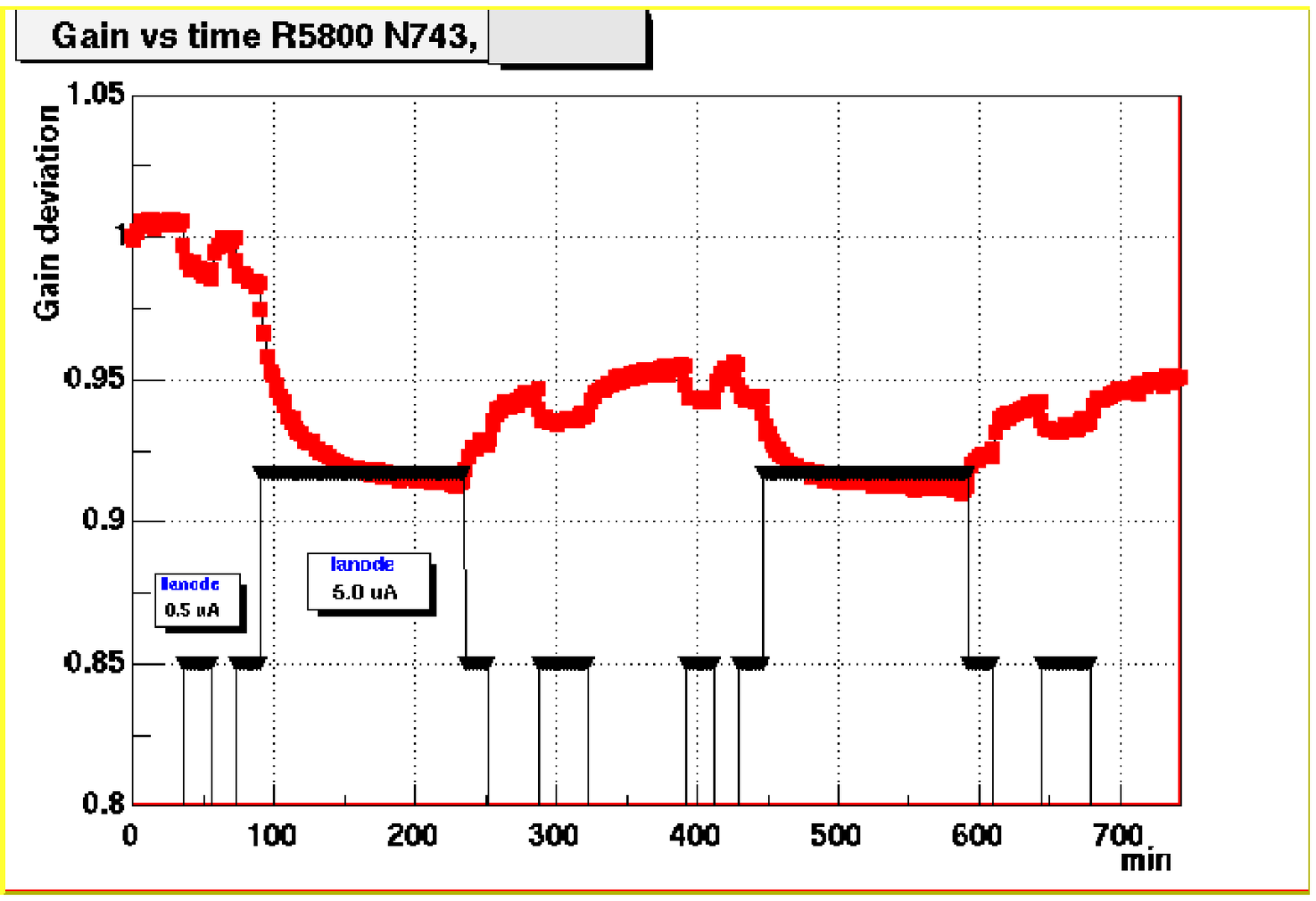,width=100mm,height=70mm}}
\caption{The behavior of PMT number 743 at the dedicated stand for PMT gain
variation measurements. (See text for details.) }
\label{fig:hm743}
\end{figure}

We used positive HV for PMT's with grounded photocathodes
for the first accelerator run and negative HV 
for the second and the third runs.
The red LED response of PMT number 743 using negative HV
during irradiation study is presented in Fig.~\ref{fig:kpm13}. The behavior of the same PMT at the
dedicated stand is shown in Fig.~\ref{fig:hm743}. We see that short-term
loss of the signal is 3-5 $\%$ when the test beam intensity is at the level 
of $10^4$ e$^-$/sec averaged over the entire accelerator cycle.

A similar signal loss was seen (Fig.~\ref{fig:hm743}) when the
additional green LED was turned on to produce the anode current of
$5~\mu$A. Another similarity is that when the ``beam'' or green LED 
was turned off, the PMT gain rose by a few 
percent. 
We compared the behavior of each PMT at the stand and during the beam test and found a satisfactory agreement between these results.\\

Six phototubes with the quartz glasses
(Hamamatsu R5800Q) of the same size were used to avoid the possible 
radiation damage to phototube windows. 
These PMT's had a gain change of 5-6$\%$ and one of them even 10$\%$. 
The gain variation of one of these quartz phototubes is presented in 
Fig.~\ref{fig:kpm13}. For the PMT's with borosilicate glasses the signal loss 
has been measured not to exceed 3$\%$ for both positive and negative HV.

\begin{figure}[hp]
\centerline{\psfig{figure= 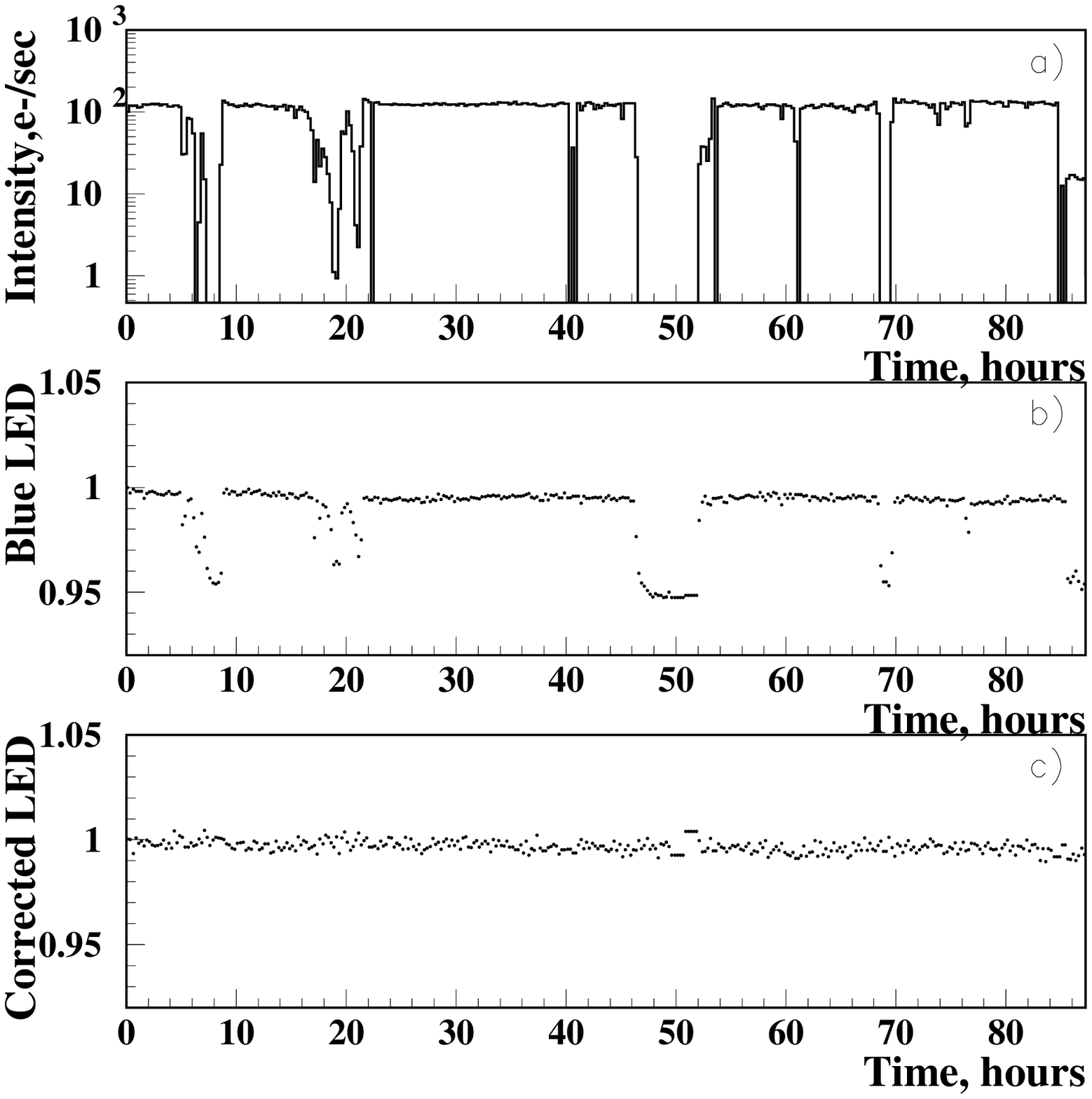,width=100mm}}
\caption{ (a) Electron beam intensity in the Shanghai crystal S22 over time. 
Blue LED time behavior
in this crystal (b) before and (c) after correction using red LED data. }
\label{fig:correct}
\end{figure}

The blue LED signal amplitude
over 85 hours is presented
in Fig.~\ref{fig:correct}. Fig.~\ref{fig:correct}(a) shows the electron beam
intensity over this time period in a sample crystal. Fig~\ref{fig:correct}(b) shows 
the raw blue LED signal for the same crystal.
We see that the blue LED signal fell by 5-6$\%$ when the beam
was off. The time diagram of
the blue LED corrected by the red LED is shown in Fig.~\ref{fig:correct}(c). 
Note that most of our PMT's lost gain when the beam was on. In our plot we 
selected this PMT with the opposite behaviour to show that we could correct 
for this  big gain change even though the sign of the change was atypical. 

All of our analyses included corrections using the red LED data. We corrected 
the signals from electrons and blue LED  on the signal from red LED 
to subtract a PMT gain variation effect from the total signal for each PMT. When 
the green or yellow LED signals were used in the analysis, they also were corrected
using the signal from red LED.

To check and correct the stability of the red LED, we used the 
$\alpha$-source. The instability of blue,
green and yellow LEDs was corrected using the PIN diode. 
The ratio of the PIN to $\alpha$ signals was stable to an accuracy 
of 0.1 $\%$. 
To decrease any possible remaining LED 
instability left after these corrections, we kept for further analysis only
accelerator spills with similar 
beam intensity.  We conservatively estimate that the error for 
the blue LED signal is 0.2$\%$.

\subsection{Irradiation by high-energy electrons}  

     The crystal array was irradiated by 27 GeV electrons 
for one week with an accelerator efficiency of 85$\%$. 
The beam intensity at the 
crystal array was 6$\times$10$^5$ particles/spill most of the time during this period.
The 80$\%$ of the beam entered in one of the six central crystals.
About a half of the time, the beam was centered on one crystal in the array and
during the rest of the time it was centered on another crystal. 
Coordinates of the electrons entering the crystal array were measured by the
drift chambers. The events with electrons near the center of the crystals
were selected for data analysis. \\

     We now describe the analysis of the electron beam data. All 
the information
which was accumulated during 85 beam hours (one position of the beam at the 
array,
see above) was divided into pieces of two hours long each. This choice was made to have enough
statistics to measure the average energy deposit in a crystal with an accuracy of
0.3$\%$, and thus we could continuously monitor the crystal signal loss. 
Prior to the irradiation study, the PMT gain of the each 
crystal in the array was adjusted to 
10,000 ADC counts when 27 GeV electron hit the center of the crystal. 
Since this corresponds to 76$\%$ of the full 
electron energy~\cite{nim2}, 
one ADC count corresponded to 2 MeV. The size of the beam spot was 
chosen 4x4 $mm^2$ for most irradiated crystals and 6x6 $mm^2$ for crystals with lower doses in order 
to equalize the statistics. The true coordinates of a particle at the array 
was calculated
with the information from the last drift chamber which was close to the
array. 
The accumulated energy peaks were fitted by a Gaussian. Then
the mean values were corrected using the red LED. \\
\begin{figure}[htp]
\centerline{\psfig{figure=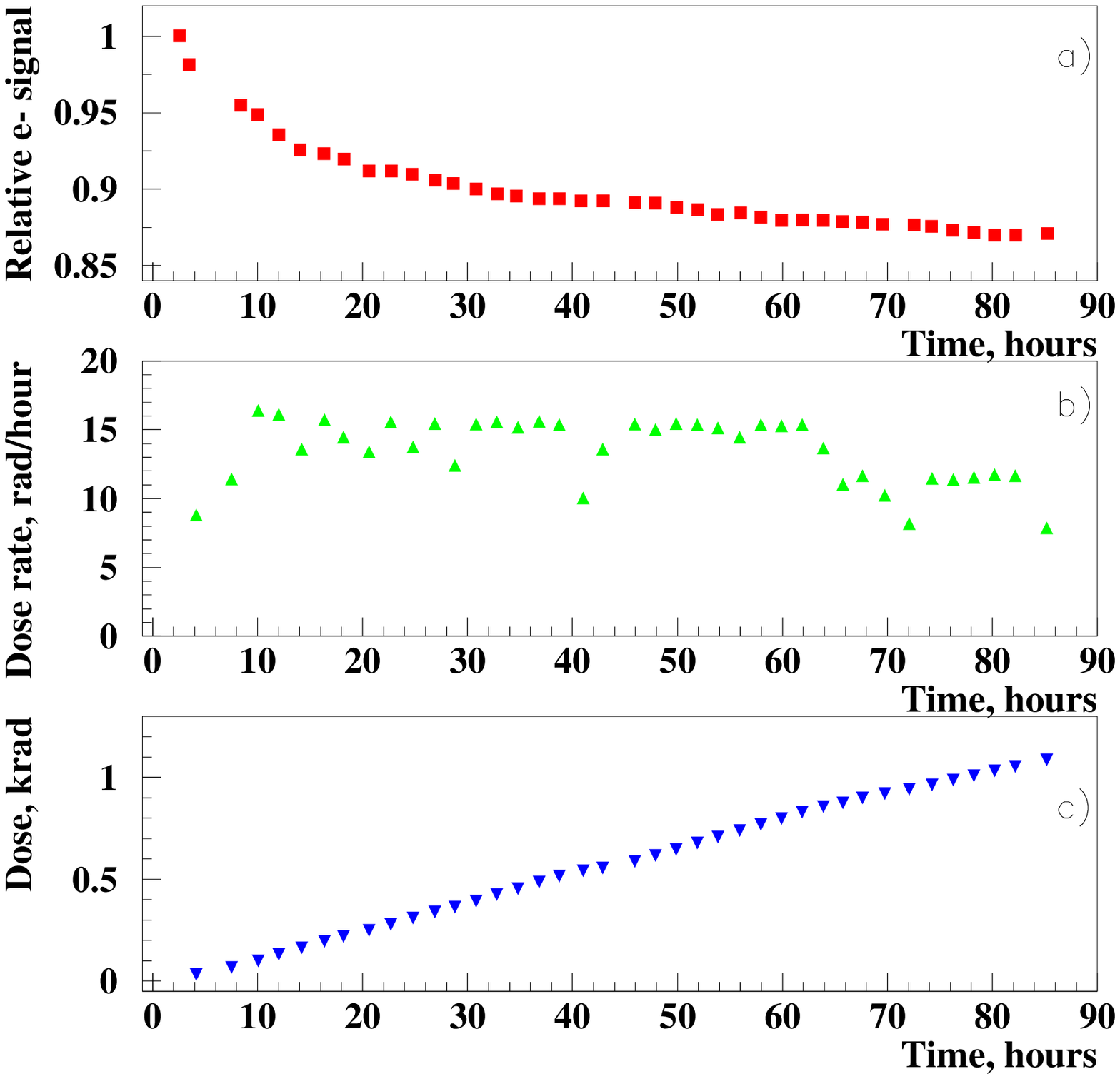,width=100mm,height=100mm}}
\caption{ (a) Normalized electron signal during 85 hours of irradiation 
by 27 GeV electrons for the Bogoroditsk crystal B14. 
(b) Electron beam intensity in dose rate units. 
(c) Absorbed dose.} 
\label{fig:apr02-1}
\end{figure}

     For each crystal a dose rate was defined as an effective number N 
of electrons per second
hitting this crystal multiplied by $25.9\cdot 10^{-4}$ 
(see Fig.~\ref{fig:btev_testbeam}(d) ).
The number N was calculated as energy deposit in this crystal in GeV/sec divided by
20.5 GeV (it corresponds to 76$\%$ of 27 GeV energy deposit when electron hits
the center of the crystal~\cite{nim2} in accordance with the MARS simulation).
  
     A typical result for an irradiated crystal is presented in Fig.~\ref{fig:apr02-1}.
In Fig.~\ref{fig:apr02-1}(b) we see an intensity of the electron beam which is 
shown in dose rate units at the shower maximum according to the MARS 
simulation results 
presented in Fig.~\ref{fig:btev_testbeam}.  The absorbed dose is given in  Fig.~\ref{fig:apr02-1}(c). The main result is shown in
Fig.~\ref{fig:apr02-1}(a) which is the normalized electron signal. We see
that finally the crystal lost 12$\%$ of the signal under an electron beam irradiation 
mostly with 15 rad/h dose rate after it accumulated 1.2 krad absorbed dose. It also appears 
that the radiation damage is saturating. 
For dose rates 
of 10-25 rad/h under 27 GeV electron beam irradiation, eight crystals
lost an average of 8$\%$  after a total accumulated dose of 1-2 krad. \\    

In order to use the light monitoring system to track the effects of 
radiation damage, it is necessary to determine the relation between the change observed by the monitoring system 
and the change in the signal from beam electrons. Because of the different optical paths 
taken by the injected monitoring light as compared to the scintillation light this constant 
is not expected to be unity. Furthermore the LED system monitors the transparency of the crystal 
at a specific wavelengths
and thus does not sample the entire spectrum of scintillation light.

\begin{figure}[ht]
\flushleft{\psfig{figure=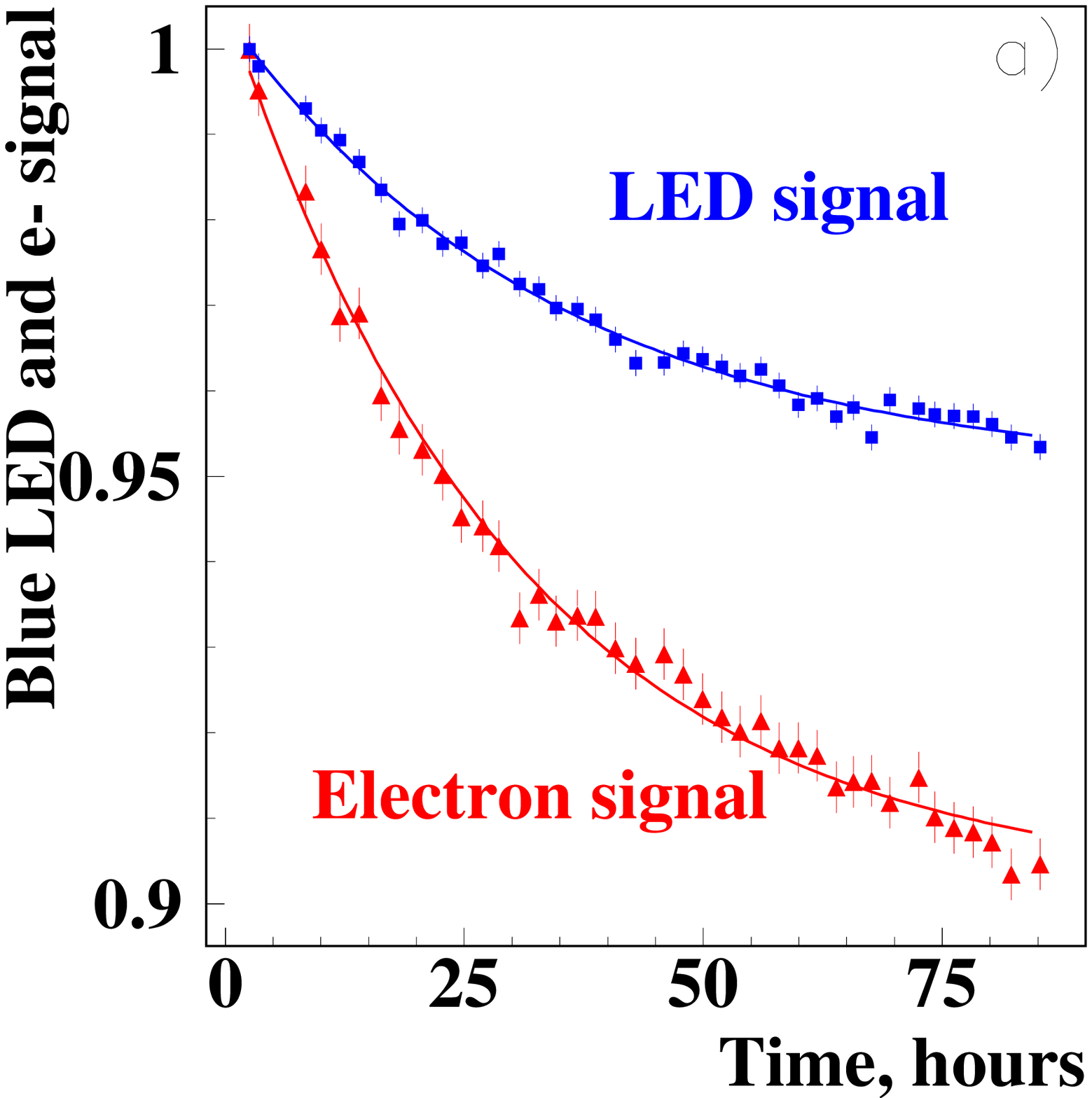,width=55mm,height=55mm}}
\vspace {-57mm}
 
\centerline{\psfig{figure=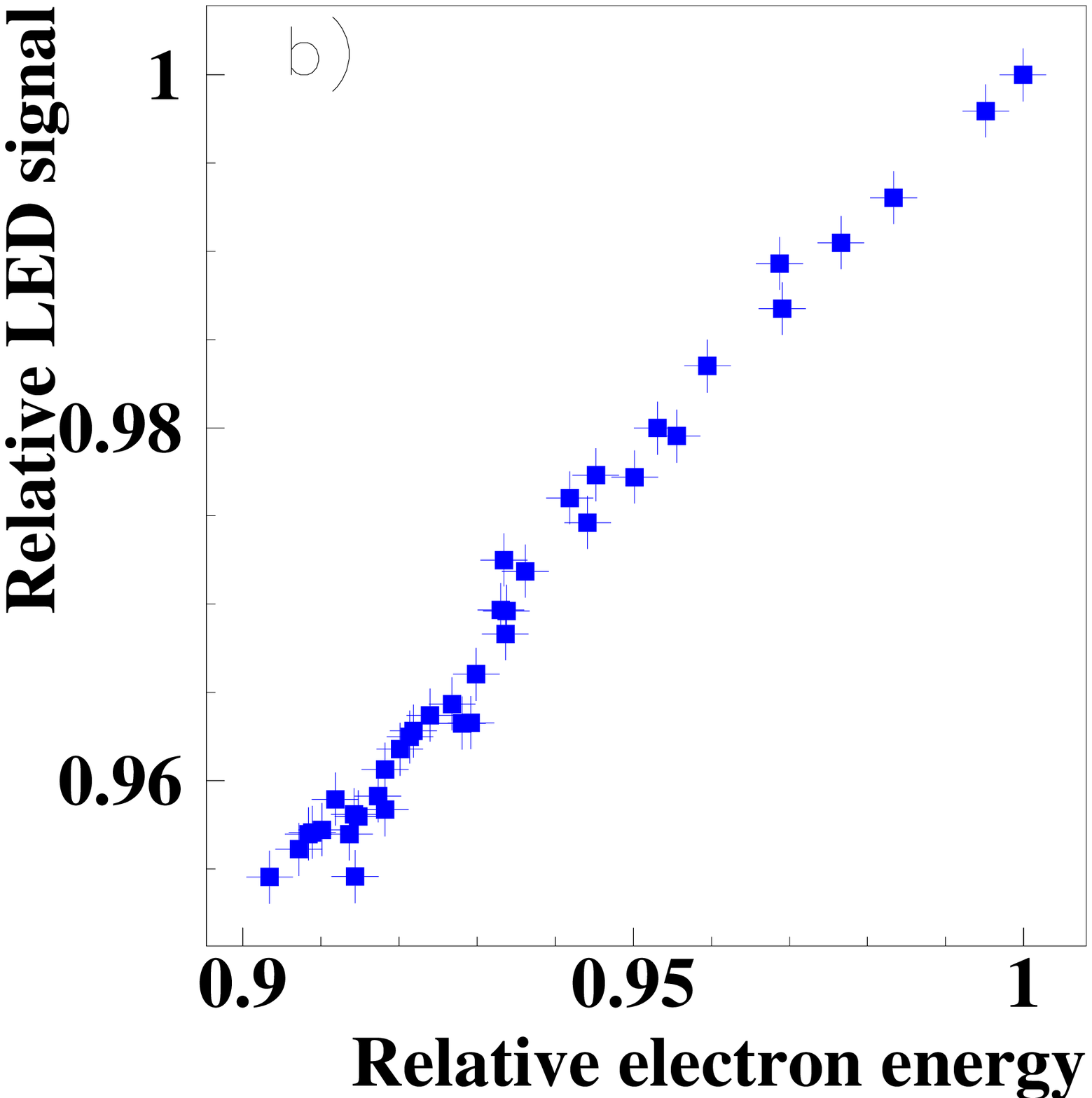,width=55mm,height=60mm}}

\vspace {-63.2mm} 
\flushright{\psfig{figure=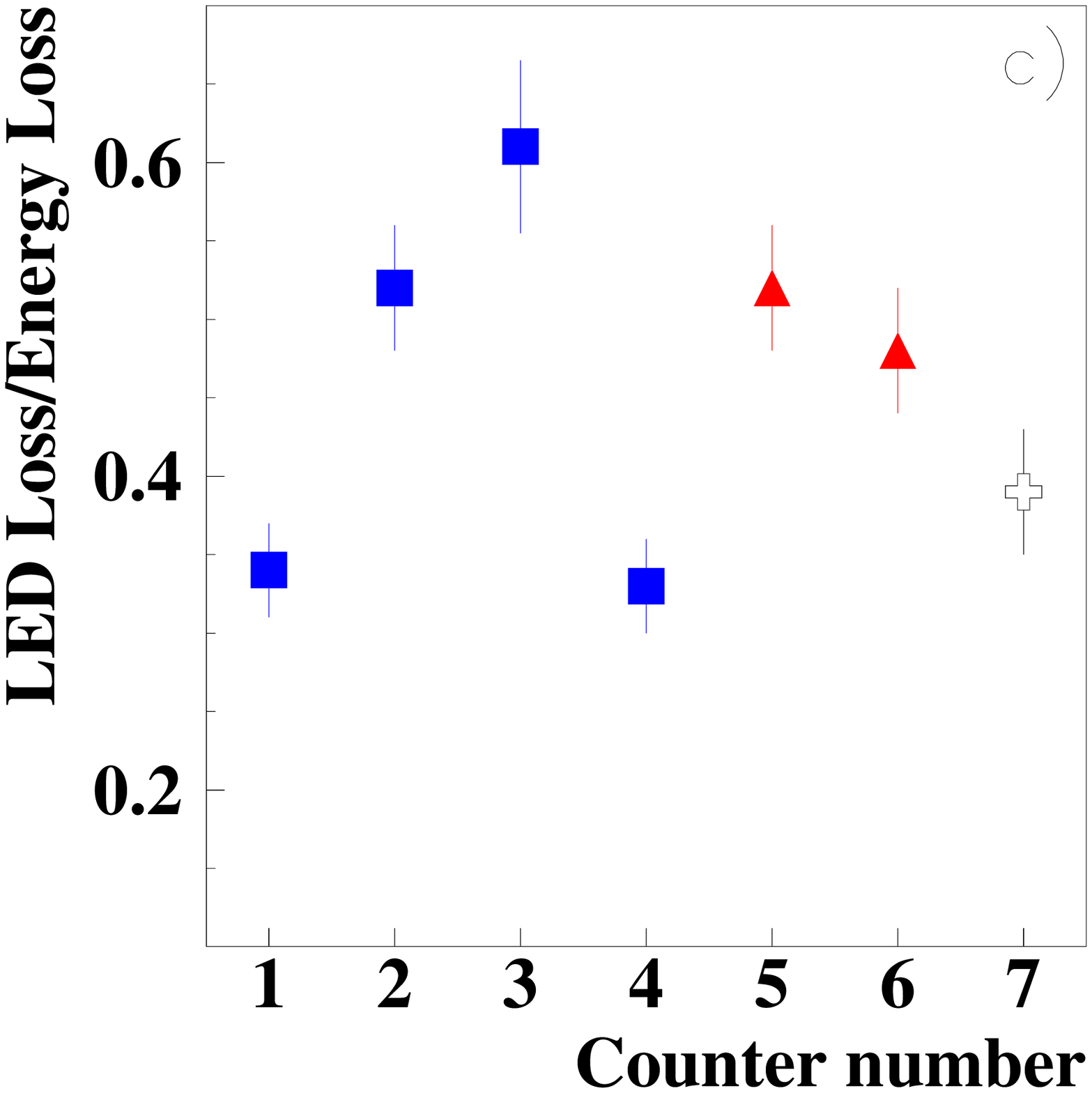,width=55mm,height=60mm}}
\caption{(a) Blue LED and electron signals for the Shanghai crystal S22, 
which was irradiated by
27 GeV electrons with a dose rate of 16 rad/h. (b) Blue LED-electron
correlation for the same crystal. (c) LED-electron correlation 
coefficients for the seven crystals. Irradiation was by 27 GeV electrons. 
Square points stand for the Bogoroditsk crystals B12, B13, B14 and B17, 
triangular points stand for the Shanghai crystals S14 and S22, and a crest 
point stands for the Apatity crystal 1447.  }
\label{fig:apr02-3a}
\end{figure}

\begin{figure}[ht]
\centerline{\psfig{figure=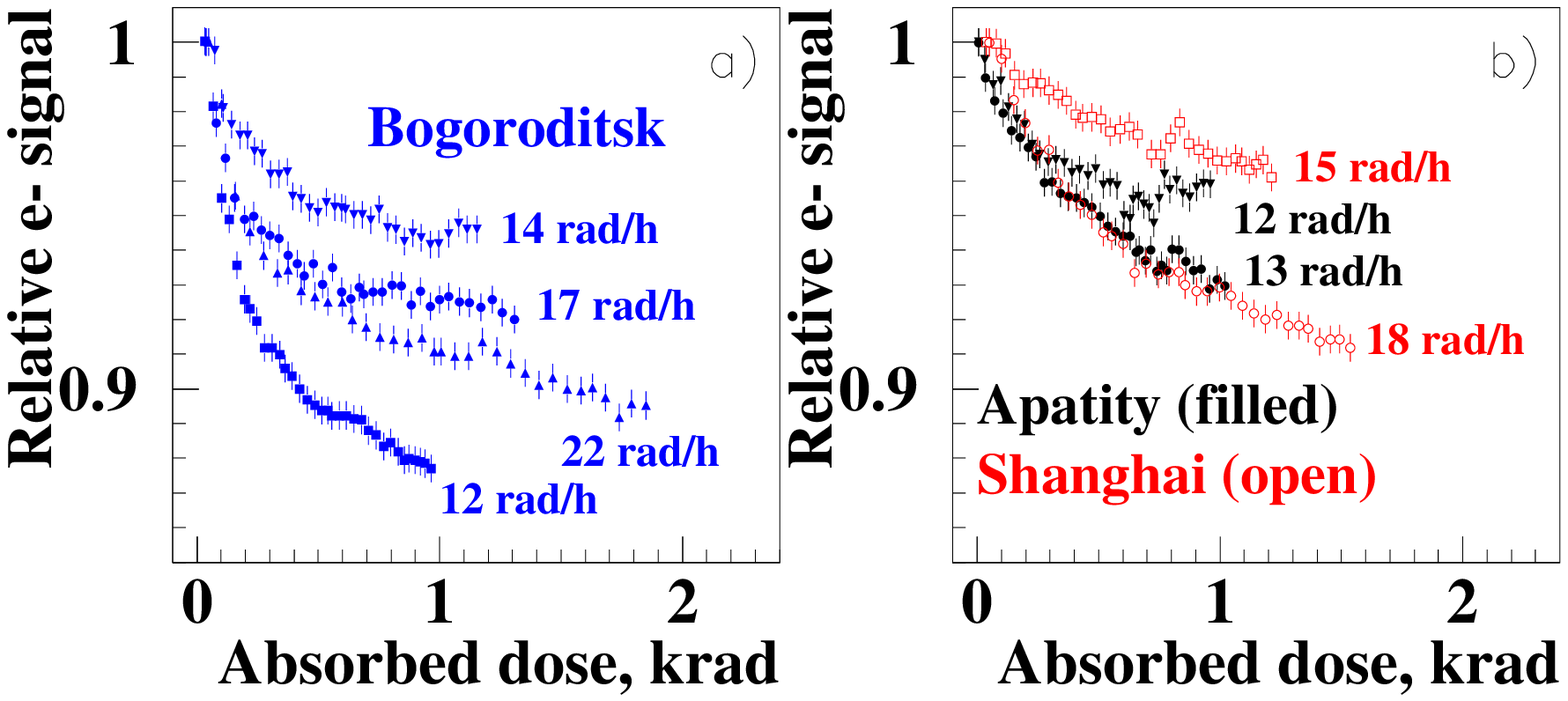,width=120mm,height=60mm}}
\caption{(a) The dependence of the electron signal on the
absorbed dose for the Bogoroditsk crystals B14, B22, B12, B16, (b) for 
the Shanghai (open points) S22, S14 and the Apatity (filled points) 1447, 1434 crystals. 
Each crystal was irradiated by 27 GeV electrons at the fixed dose rate(different
for each crystal) for 85 hours.}
\label{fig:apr02-4}
\end{figure}

The blue LED emits at 470 nm and the scintillation peak 
is at 430 nm. The typical blue LED and electron signal behavior under irradiation for one of the 
crystals is shown in Fig.~\ref{fig:apr02-3a}(a). 
The blue LED (as well as the electron signal) is corrected by the red LED, and the red LED
by the  $\alpha$-source. The same was provided for the green and yellow LEDs. 
For the green LED a signal loss was smaller than for the blue LED, and for the
yellow LED the signal loss was smaller still (not shown). In 
Fig.~\ref{fig:apr02-3a}(b) we see a strong correlation between the change in the blue LED 
light level and the beam signal. 
We fit such distributions by the
straight lines, ignoring some deviations from linearity.
The results for a few crystals are presented in Fig.~\ref{fig:apr02-3a}(c). 
We did not observe 
a significant difference in the crystals from different manufacturers. Constants of 
proportionality vary from 0.3 to 0.6 for these 
crystals. The dependence  of a relative electron signal on the
absorbed dose is presented in Fig.~\ref{fig:apr02-4}. \\

 A simple model is used to describe  signal loss. The signal loss $dy$ is
proportional to the signal value $y$ and the  number of the produced color centers, 
which are proportional to the absorbed dose $dR$. Crystal recovery is proportional to a
difference between the asymptotic value $y_0$(after recovery) and the current signal value. 
Also it is proportional to the recovery time $dt$:
\begin{equation}
 dy = -P1\cdot ydR + P2(y_0 - y)dt = (-(P1\frac{dR}{dt}+P2)y+P2\cdot y_0)dt 
\end{equation}

In our case the dose rate($\frac{dR}{dt}$) was almost the same during the 85 hours 
of irradiation. 
Integration of this equation gives us the expression:

\begin{equation}
y=P0\cdot \exp^{-(P1 \frac{dR}{dt}+P2) \cdot t } + \frac {P2\cdot y_0}{P1 \frac{dR}{dt}+P2} 
\end{equation}

We can
present the signal loss behaviour function as  
\begin{equation} 
f(t) = a \cdot \exp^{ -t/ \tau} + (1-a),
\end{equation}
The results of the fit for Fig.~\ref{fig:apr02-3a}(a)  are 
listed in Table~\ref{tab:expon}.

\begin{table}[hbt]
\begin{center}
\caption{Results of fits to $ f(t) = a \cdot \exp^{ -t/ \tau} + (1-a)$ }
\label{tab:expon}
\begin{tabular}{cccc}\hline 
Signal Source & $a$ & $\tau,hour$  \\\hline
Electron beam & 0.104$\pm$0.002 &  30$\pm$2 \\
Blue LED & 0.054$\pm$0.002 & 34$\pm$5 \\
\\\hline
\end{tabular}
\end{center}
\end{table}

The parameter $a$ defines the saturated light loss value that is reached as $t$ goes to
infinity at a constant dose rate. 
Close to the asymptotic value, the crystal lost 10$\%$ in the electron 
signal and 5$\%$ in the blue LED signal. The $\tau$ parameter defines 
the saturation time constant, which  is 30 hours for our crystal and our
dose rate. 

The time constants for the ten studied crystals are between 20 and 30 hours. 
There is no significant difference in $\tau$ for the LED and 
electron signals.\\

We should make a note at the end of this section.
When a crystal is irradiated, the red LED light is slightly
absorbed. Herewith, the blue LED light is absorbed more,
in 3-6 times more~\cite{red_led} compare to the red LED light
in the crystals.
We can estimate that the electron signal is absorbed in about
two times more than the blue LED signal (see Fig.~\ref{fig:apr02-3a}(c)).
We assumed that red light was unchanged under crystal
irradiation, and assigned the PMT gain change to the red
LED change. It means that the absolute electron signal loss
values might be in about 1.1 times higher than the
presented ones.
 
\subsection{Irradiation by high-energy pions}  

After the electron irradiation program was finished, we irradiated 
the same crystals with pions for a four day period.
We used a 40 GeV $\pi^-$ beam. 
The size of the 40 GeV pion beam was 8 cm horizontally and 6 cm vertically, 
$i. e.$ 90\% of the beam was contained within these dimensions. 
The beam intensity was 6 $\times 10^6$/sec. Six crystals
were irradiated with a dose rate ranging from 10 to 30 rad/h. Five
cycles of irradiation (15-20 hours each) were alternated by low intensity
electron beam exposures to measure the scintillation signals in the crystals.

\begin{figure}[ht]
\centerline{\psfig{figure=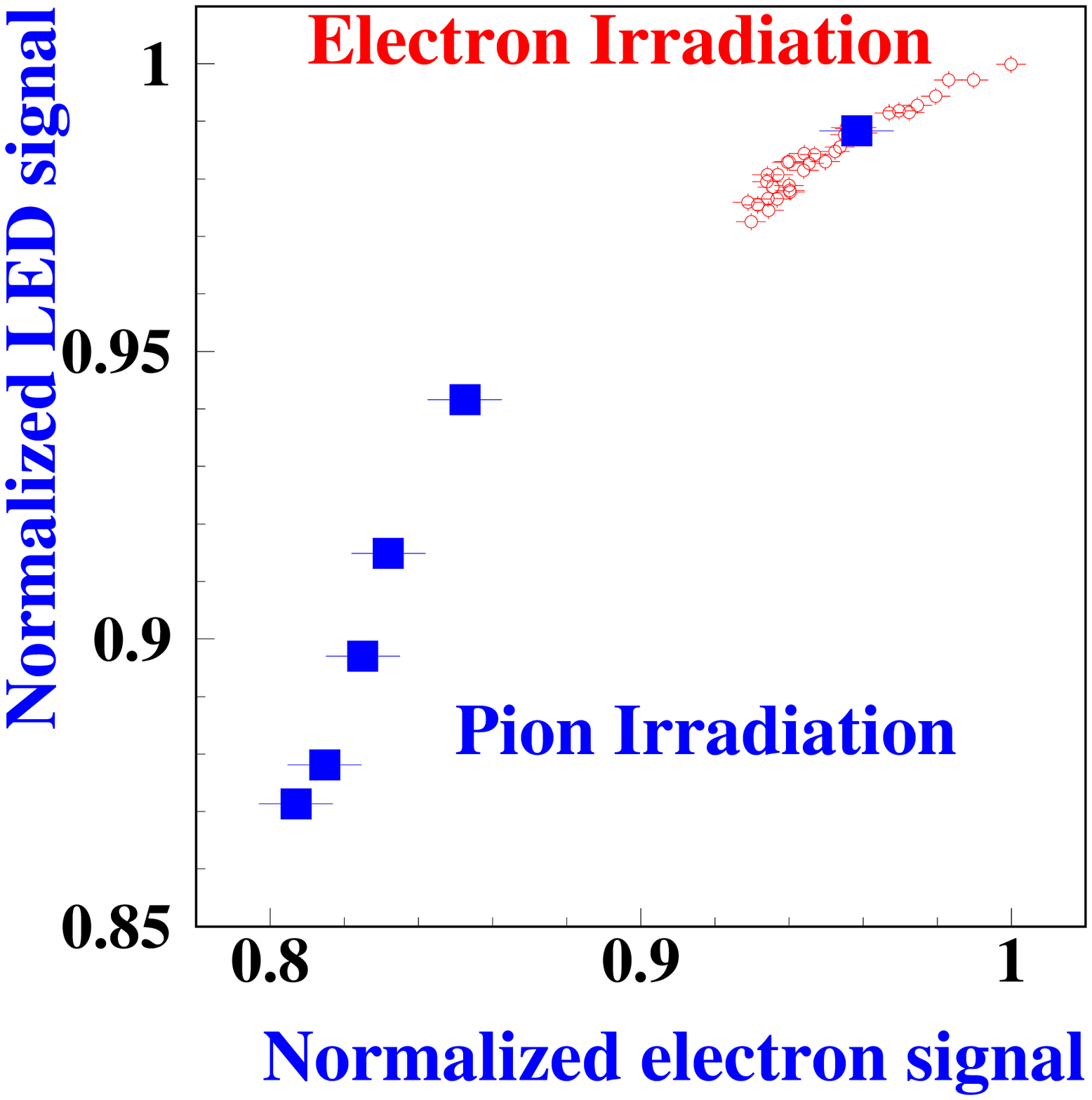,width=85mm,height=85mm}}
\caption{ Normalized LED signal versus normalized electron signal for Apatity
crystal 1447. Open points show electron irradiation and filled square points show
further pion irradiation. }
\label{fig:cell14_pi}
\end{figure}

The radiation damage region in the crystals is different for an electron
and a pion irradiation (see Fig.~\ref{fig:btev_testbeam}). Thus, if a crystal
was irradiated first by electrons until saturation in radiation damage was reached for a given 
dose rate, then 
we expect to get an additional signal loss with pion irradiation even at the same dose rate.
Fig.~\ref{fig:cell14_pi} shows the additional loss of signal for one of the crystals (from Apatity). 
This crystal was irradiated
by 27 GeV electrons and then by 40 GeV pions. During the 85 hours of e$^-$ irradiation the dose
rate was 12 rad/h. Then in the next 85 hours the dose rate was an order of magnitude less, and 
the crystal recovered. As a result, the first filled square point for pion irradiation data 
is above many open points for electrons. After that the crystal was irradiated 
by pions with the dose rate 12 rad/h for 100 hours. We see that the crystal lost 8$\%$ of 
the signal during the electron irradiation period and 14$\%$ of the signal during the pion
irradiation period with the same dose rate. The constant of proportionality between the blue LED and the 
electron signal is 0.3 for electron irradiation
(if one fits by a straight line). This is about the same for pions at the beginning 
of the pion irradiation, but then increases up to 1 during the further pion 
irradiation. The crystals
(manufactured in Bogoroditsk and Apatity) lost about 14$\%$ on pion 
irradiation.
The surrounding crystals which were
irradiated with a dose rate of about 1 rad/h lost less than 1 $\%$ of
their light output.\\   

 The dependence of a signal loss on dose rate was studied in a separate
run using 40 GeV pion irradiation.
Each beam exposure lasted 
for 6 continuous hours. The beam intensity started from 2$\times 10^5$/sec
and was increased in a few steps up to 8$\times 10^6$/sec by the end of
the study. 
The beam was present in 1 sec of the full accelerator cycle of 9 sec. 
After each 6 hour irradiation exposure we lowered intensity by a few orders of
magnitude, down to 3$\times 10^4$/sec, so that we could avoid pile-up and see 
a minimum ionizing peak (MIP) for pions traversing the crystals without interacting. The crystals 
light output signals were monitored using the MIP
peak; this procedure took 2 hours at low intensity.
After that we took again high intensity beam exposure for the next 6 hours to
continue irradiating the crystals. Then again switched to the low intensity
MIP exposure. 

\begin{figure}[ht]
\flushleft{\psfig{figure=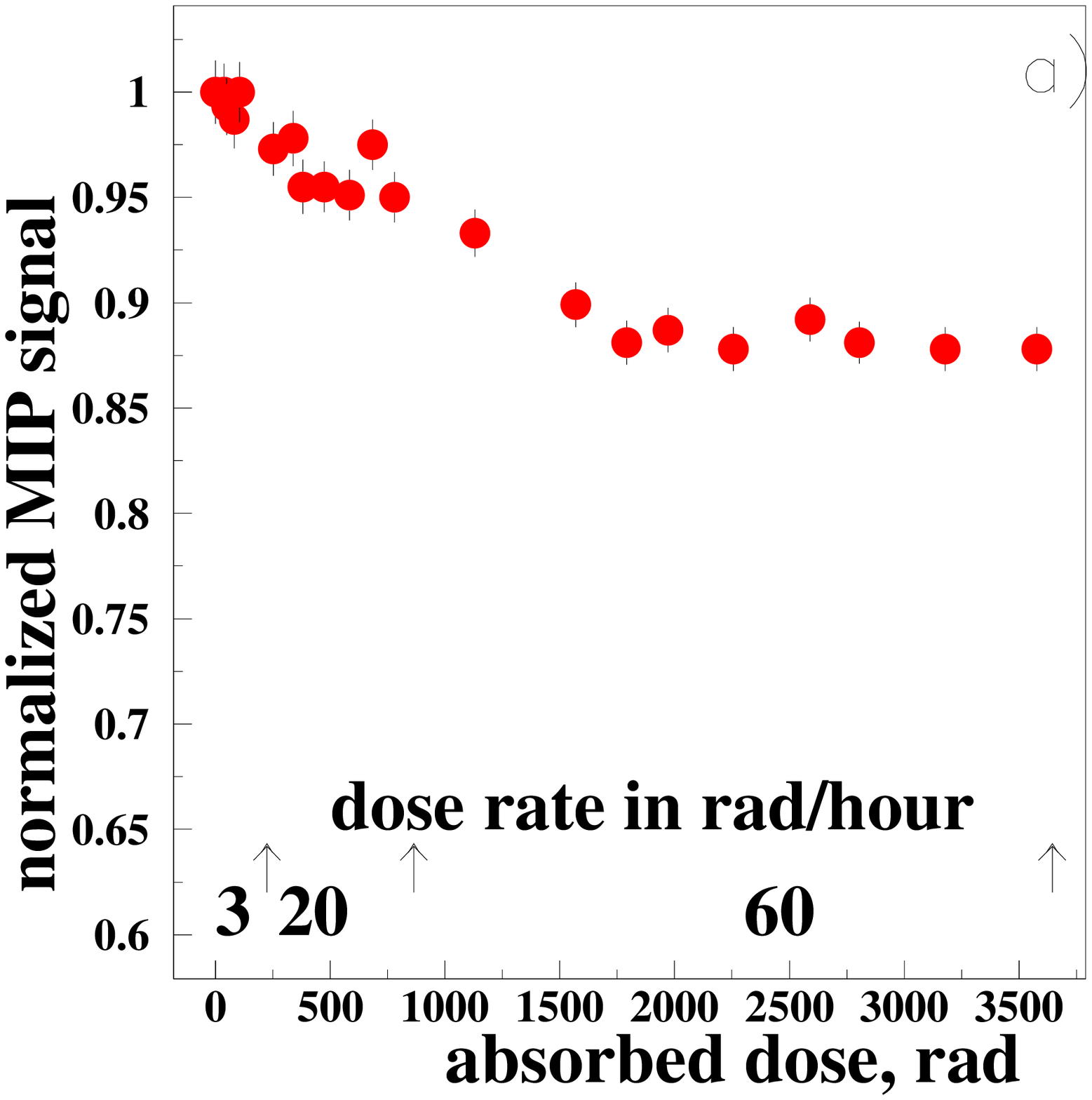,width=58mm,height=55mm}}
\vspace {-57mm}
 
\centerline{\psfig{figure=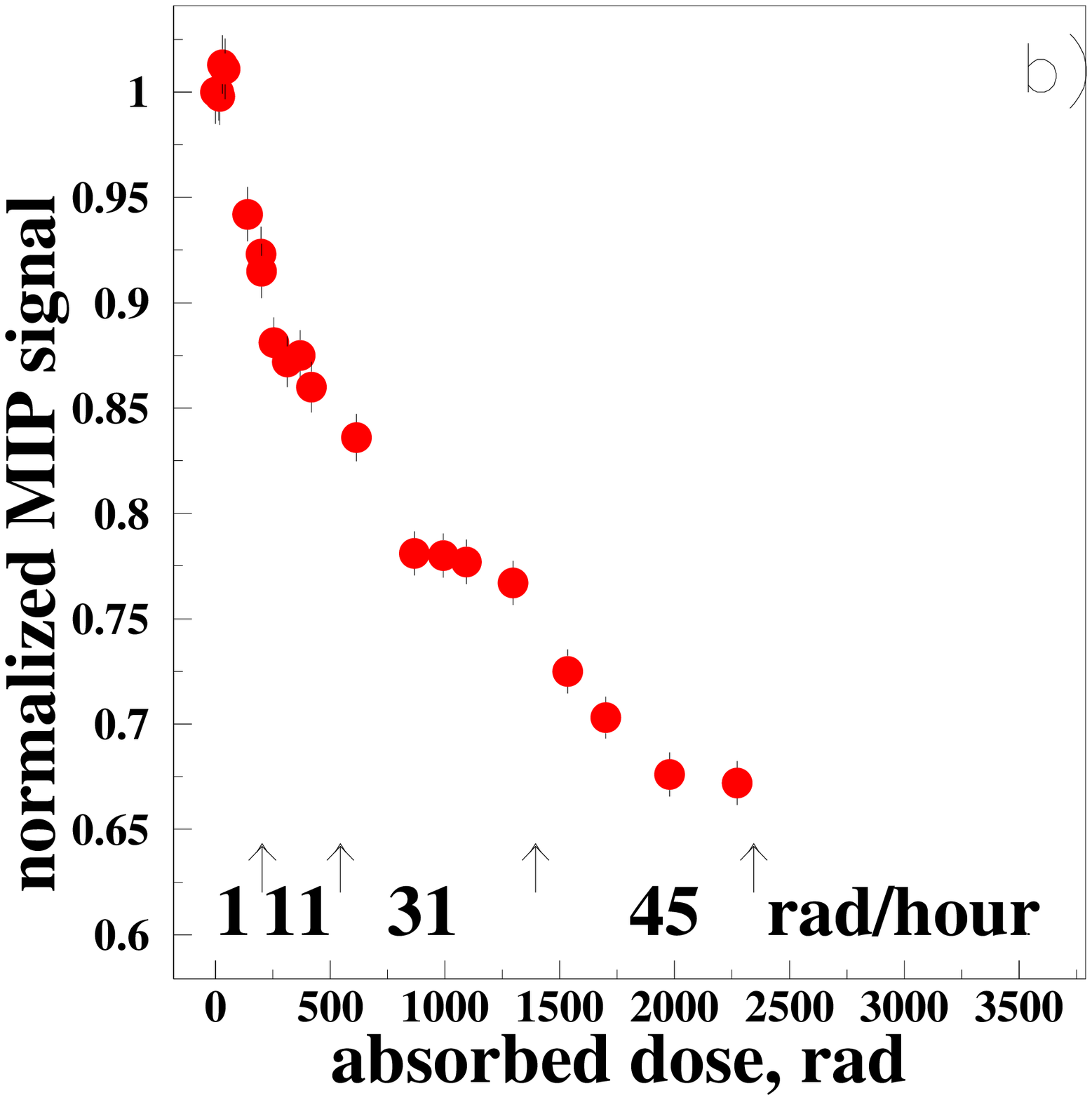,width=58mm,height=60mm}}

\vspace {-63.2mm} 
\flushright{\psfig{figure=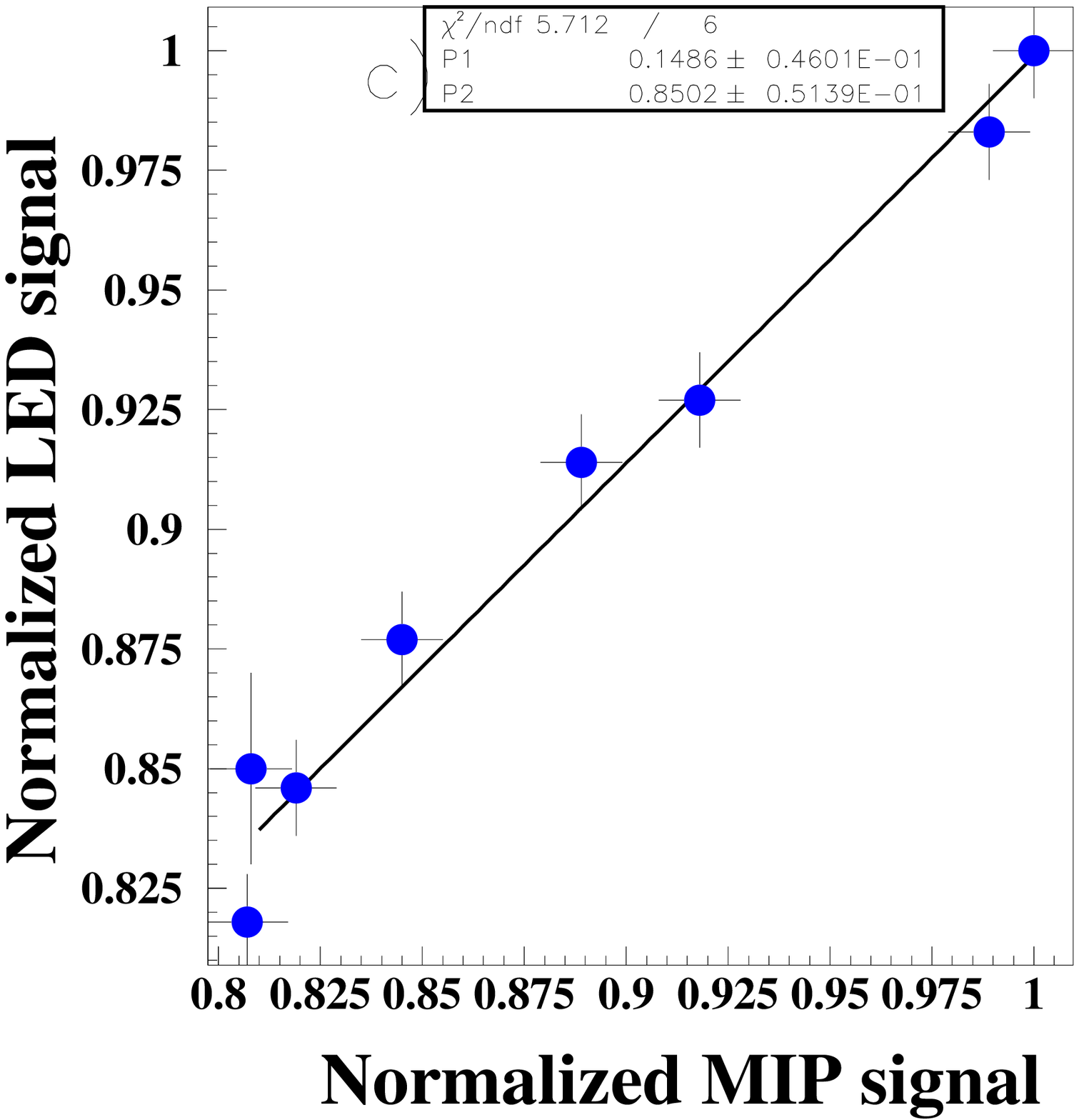,width=58mm,height=60mm}}
\caption{Dependence of the normalized MIP signal on absorbed dose for 
(a) Shanghai crystals S16 and (b) S20. (c) Correlation between 
the LED and the MIP signals for Shanghai crystal S19 as a result 
of a pion irradiation.}
\label{fig:mip_22}
\end{figure}

To check our procedure for obtaining the change in scintillation light from time to time, 
we used pure muon beams and 27 GeV electrons to measure the light output changes
due to pion irradiation.
We continued this procedure of alternating high intensity and
low intensity beams for 10 days in a row. The dependence of the normalized MIP signal on an 
absorbed dose for the two crystals in the array is shown
in Figs.~\ref{fig:mip_22}(a) and (b). (The normalized
MIP signal is defined as the ratio of the MIP signal after some absorbed
dose to the one before the pion irradiation). 

\begin{figure}[hb]
\flushleft{\psfig{figure=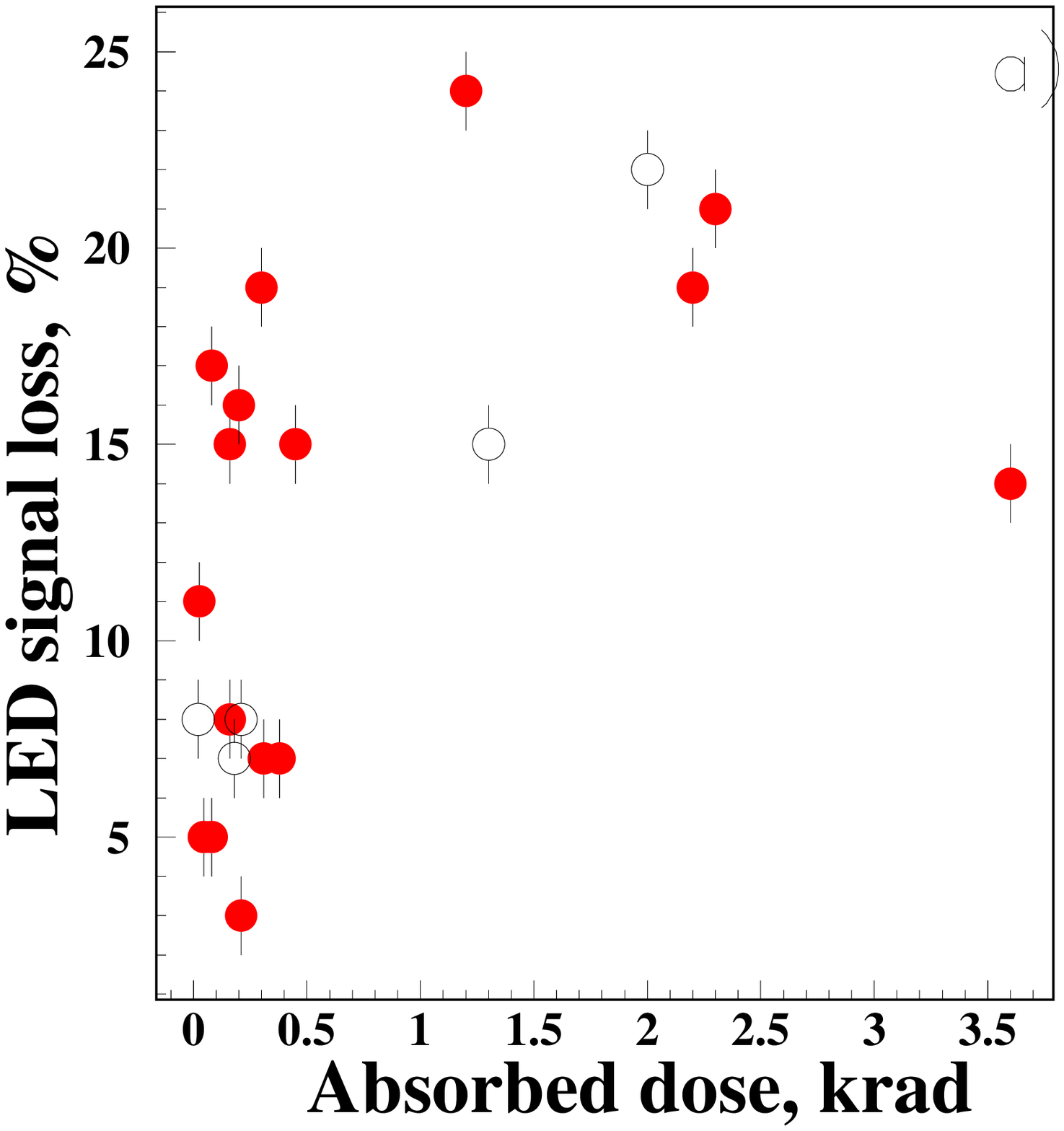,width=85mm,height=70mm}}
\vspace {-73mm} 
\flushright{\psfig{figure=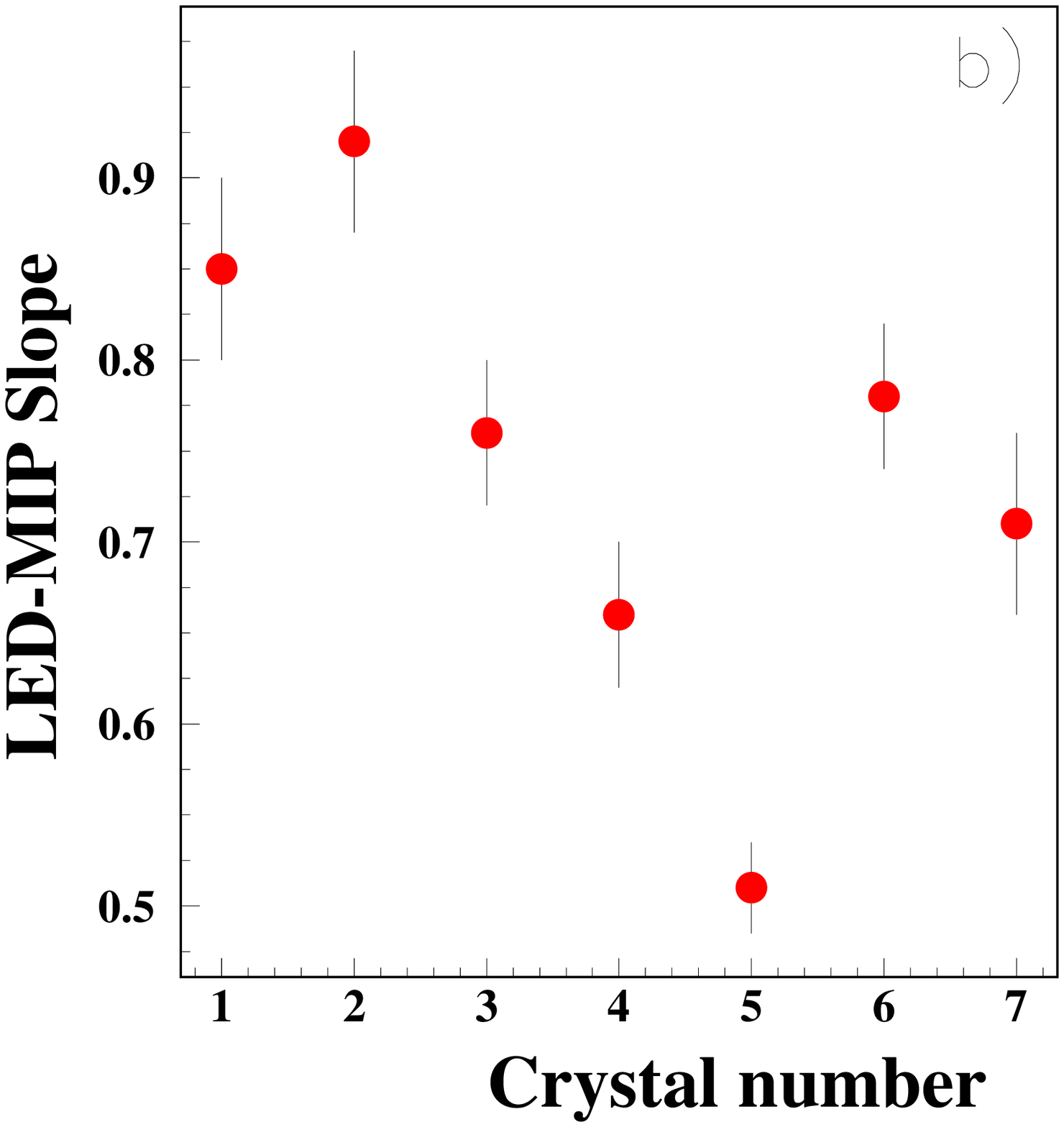,width=85mm,height=70mm}}
\caption{ (a) Dependence of a LED signal loss on 
absorbed doses obtained for a 10 day irradiation by a 40 GeV pion beam of the
21 crystals. (100 rad = 1 Gy). (b) The constants of proportionality 
between the change observed by the LED monitoring system and the change in the 
beam (MIP) signal for seven crystals. Irradiation was by 40 GeV pions. }
\label{fig:mip_led_23}
\end{figure}

We have observed the dependence of light output loss 
on the dose rate. Like electron radiation, the light loss exhibits saturation effect 
when the dose was kept at a constant level. 
The correlation between a change in
the LED signal and a change in the MIP signal under irradiation was also measured (see Fig.~\ref{fig:mip_22}(c) as an example).
The constant
of proportionality, if one fits by a straight line is different for 
different crystals and is on average 0.7
(the LED signal decreases less than the MIP one). 
In Fig.~\ref{fig:mip_led_23}(a) we show the decrease in the LED signal for moderate dose 
rates.  
Different crystals received different absorbed doses during 10-day irradiation period.
The open circles stand for 
Bogoroditsk crystals and the filled circles stand for Shanghai
crystals.  Six points in Fig.~\ref{fig:mip_led_23}(a) represent the six crystals 
described above which accumulated 
absorbed doses of more than 1 krad each. Fifteen other crystals were irradiated
by the beam halo and received absorbed doses less than 500 rad each. They
are shown on the left side  of Fig.~\ref{fig:mip_led_23}(a) . Three Bogoroditsk crystals 
are close to
each other in their radiation hardness, however the twelve Shanghai crystals 
differ among each other by an order of magnitude. \\

     Irradiation of lead tungstate crystals creates color centers 
which reduce the light attenuation length. One expects that the change of 
attenuation length will affect the longitudinal uniformity. This can degrade the energy 
resolution. On the other hand, if the loss of light collected in the crystal after 
irradiation is relatively small, the energy resolution itself might not be 
degraded so that the radiation damage can be regarded as only 
a calibration issue. 
 The non-uniformity of the light yield (LY) along the crystal contributes to
the energy resolution. To measure changes in the LY non-uniformity the
crystal array was rotated by 90 $^0$ with respect to the beam direction,
before the irradiation by pions and just after the 10 days of irradiation. 
The crystals were scanned with the muon beam. 
The position of the muon track
going through the crystal was reconstructed with the drift chambers. The data
were binned along the crystal lengths in 1 cm intervals. The energy deposit 
distribution was fitted in each bin by a convolution of Gaussian and Landau function.   
The non-uniformity of the light yield in the front part of
crystal (3-10 radiation lengths) was about 0.5 $\%$/cm. The non-uniformity did not
change significantly after a dose up to 4 krad at a
dose rate of up to 60 rad/h, which caused the signal loss of up to
30$\%$. As a result, the energy resolution of the crystals did not
change.  
The relation between the change in transparency seen by
the LED light and the change seen by the scintillation light varies from
crystal to crystal. A plot of such constants of proportionality for
seven crystals is shown in the Fig.~\ref{fig:mip_led_23}(b). 
The first four points show the Shanghai crystals, and the next three points 
show the Bogoroditsk crystals. Points 4 and 5 represent the 
super-intensive dose rates obtained by the Shanghai crystal S25 and 
the Bogoroditsk crystal B21 (details will be given in the next Section).
We can see that the constants of proportionality for 40 GeV pion
irradiation (Fig.~\ref{fig:mip_led_23}(b) ) are larger
than the constants of proportionality for 27 GeV electron irradiation
(Fig.~\ref{fig:apr02-3a}(c) ).\\

\begin{figure}[h]
\centerline{\psfig{figure=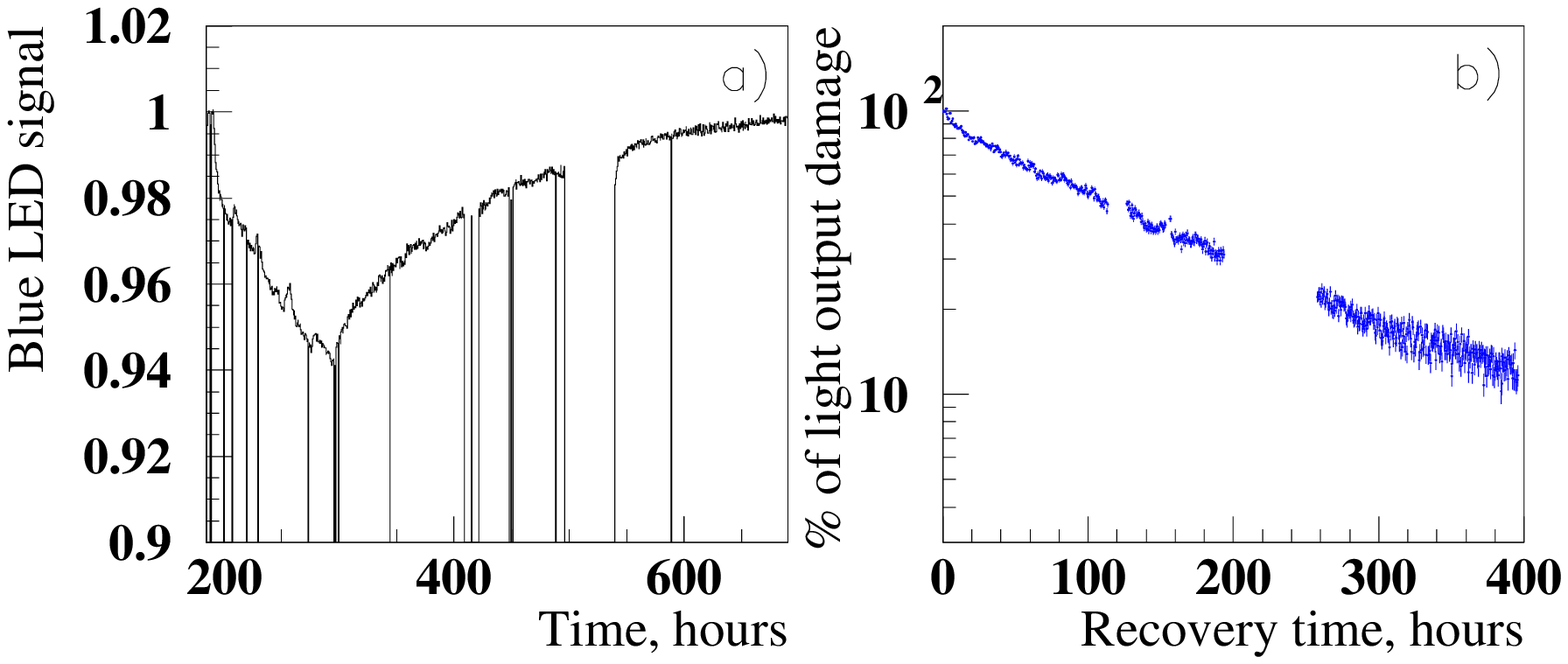,width=160mm,height=70mm}}
\caption{ (a) Blue LED signal of the Apatity 1434 crystal which was irradiated
by 40 GeV pions during 100 hours and then recovered during the next 
400 hours. (b) The blue LED light output damage recovery in the 
same crysral. 
We see that the crystal recovered 90$\%$ of 
its light output damage after 400 hours. }
\label{fig:recovery}
\end{figure}

After the irradiation by electrons and then pions was finished, we kept the
PMT HV on and studied crystal recovery for 15 days using the LED pulser. 
The results for the Apatity 1434 crystal are presented in Fig.~\ref{fig:recovery}. We 
fitted the dependences of recovery on time for the six crystals with an exponential 
function. The average recovery time is (200$\pm$40) hours, and the LED damage recovery
for 400 hours is (87$\pm$5)$\%$ for these six crystals. \\   

\section{Super-intensive beam irradiation}

\begin{figure}[tb]
\centerline{\psfig{figure=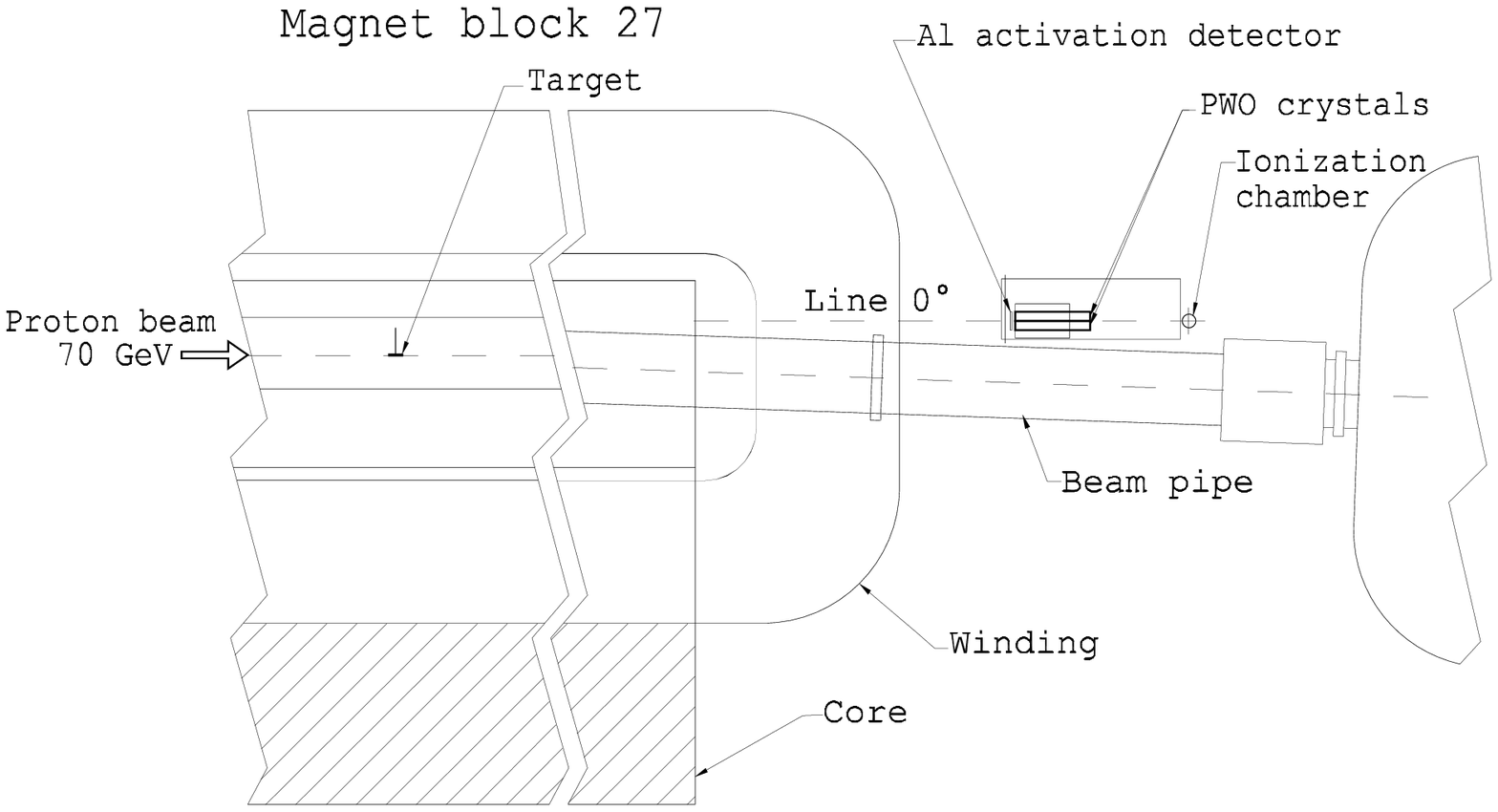,width=100mm,height=75mm}}
\caption{Superintensive dose irradiation facility}
\label{fig:block27}
\end{figure}

Six crystals from Bogoroditsk and Shanghai  
were irradiated by secondary particles coming out the internal target of
the 27-th magnet block of the Protvino U-70 
accelerator(see Fig.\ref{fig:block27}). Two of them were irradiated
at a dose rate of 100 krad/h, and the other four at 1 krad/h.
For the latter case the intensity of the
primary proton beam was lowered
by two orders of magnitude. To measure the absorbed dose, thermo-luminescence 
dosimeters (TLD)
were attached to the front face of the crystals.
 They were of LiF type doped by Mg, Cu and P, 5 mm in diameter and 
200 $\mu$m in thickness. In addition, an ionization 
chamber(IC) filled by Xenon was installed behind the crystals. 
The sensitive volume of
the chamber was as 18.5 mm in diameter and 36 mm in length.
Both TLDs and the IC were calibrated using
a Cs-137 gamma source. The accuracy of the absorbed dose measurements
 by TLDs and IC in this mixed radiation field
was estimated to be $30\%$ each. These measurements were in general
agreement with the results of the MARS calculations; the worst case difference
was a factor of 1.5. The dominant systematic error 
of the calculations was due to the
accuracy of the irradiation facility geometry.
  The IC was used to monitor the number of protons produced 
 at the internal target for each run of the crystal irradiation.
$Al$ activation detector in 
Fig.\ref{fig:block27} was used to measure a fluence of the hadrons 
(number of hadrons per cm$^2$) with energy greater than 20 MeV.\\

Two crystals, Bogoroditsk B21 and Shanghai S25, were irradiated
in the first exposure at about 100 krad/h dose rate.
 The longitudinal profiles of the absorbed dose rates are shown 
in Fig.~\ref{fig:crazy_profiles}(a).
The maximal values of the absorbed doses accumulated in crystals
during the five exposures are given in Table \ref{tab:summary1}.
Four  crystals, Bogoroditsk B17, B9 and Shanghai S22, S18,
 were irradiated in the second exposure at 1 krad/h dose rate.
 The longitudinal profiles of the absorbed dose rates are shown 
in Fig.~\ref{fig:crazy_profiles}(b).
The absorbed doses accumulated in the crystals
during the five exposures are given in Table \ref{tab:summary2}. \\
 
\begin{table}[htb]
    \caption{The maximal values of the absorbed doses accumulated
      in Bogoroditsk B21 and Shanghai S25  crystals
      during the five exposures at the IHEP irradiation facility}
    \label{tab:summary1}
    \begin{center}
      \begin{tabular}{|c|c|c|} \hline
Exposure    & Bogoroditsk B21 & Shanghai S25  \\
(minutes)   & (krad)            &  (krad)    \\ \hline \hline

0.83         & 3.4     &  1.8    \\ \hline
9.67       & 40    &  21    \\ \hline
66       & 270    &  140    \\ \hline
475         & 1970     &  1020    \\ \hline
747        & 3100    &  1610    \\ \hline
      \end{tabular}
    \end{center}
\end{table}
 
\begin{table}[htb]
    \caption{The maximal values of the absorbed doses accumulated
      in Bogoroditsk B17, B9 and Shanghai S22, S18  crystals
      during the four exposures at the IHEP irradiation facility
}
    \label{tab:summary2}
    \begin{center}
      \begin{tabular}{|c|c|c|c|c|} \hline
Exposure    & Bogor. B17 & Bogor. B9 & Shanghai S22& Shanghai S18  \\
(minutes)   & (krad)       &  (krad)     &    (krad)  &     (krad)  \\ \hline \hline

25         & 0.7     &  0.7    & 0.35    &   0.35    \\ \hline
72         & 2    &  2   & 1     & 1       \\ \hline
60         & 1.7    &  1.7   &  0.8     &  0.8       \\ \hline
60         & 1.7    &  1.7   & 0.8      &  0.8       \\ \hline
      \end{tabular}
    \end{center}
\end{table}

\begin{figure}[ph]
\flushleft{\psfig{figure=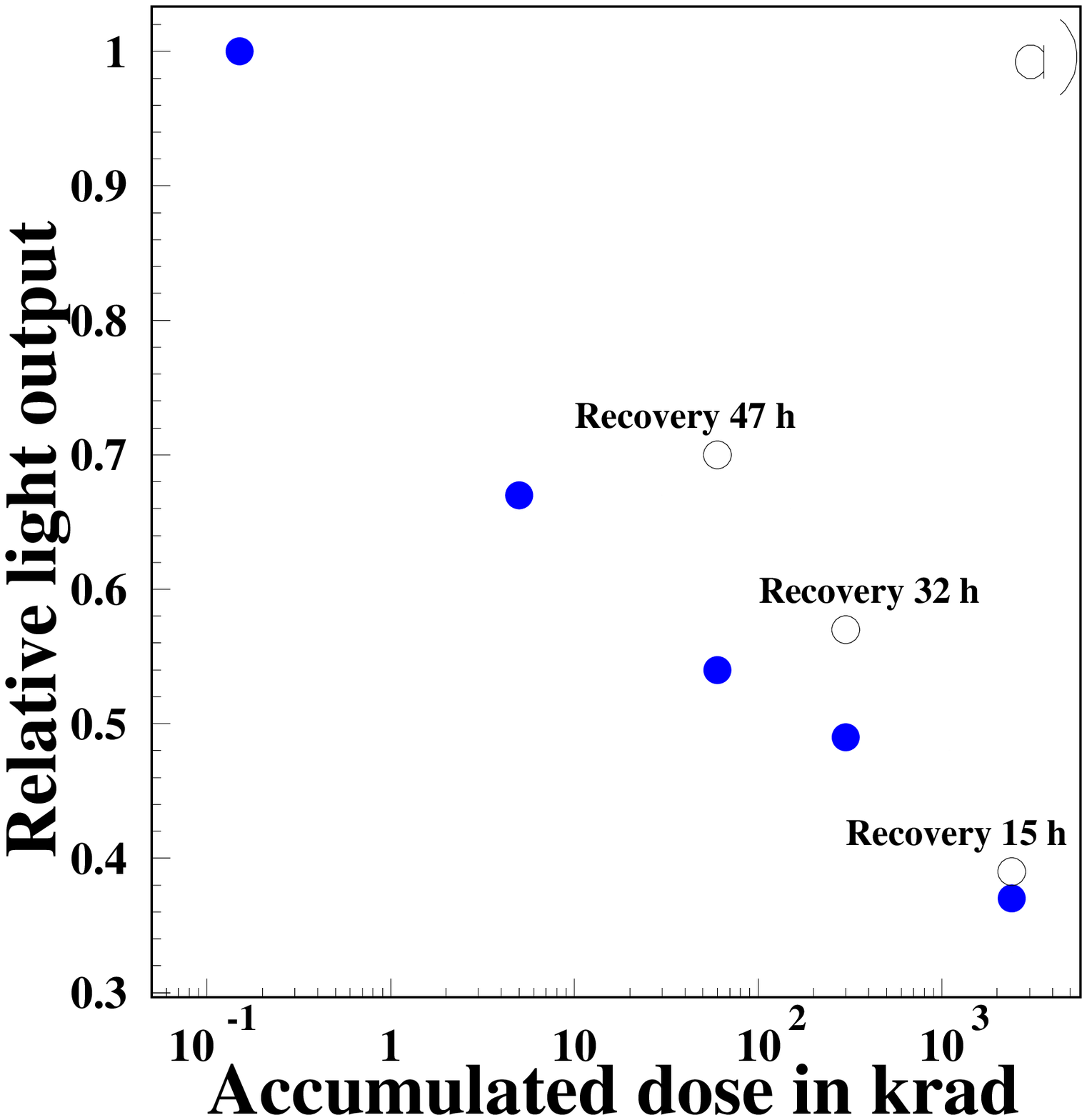,width=85mm,height=70mm}}
\vspace {-73mm}
\flushright{\psfig{figure=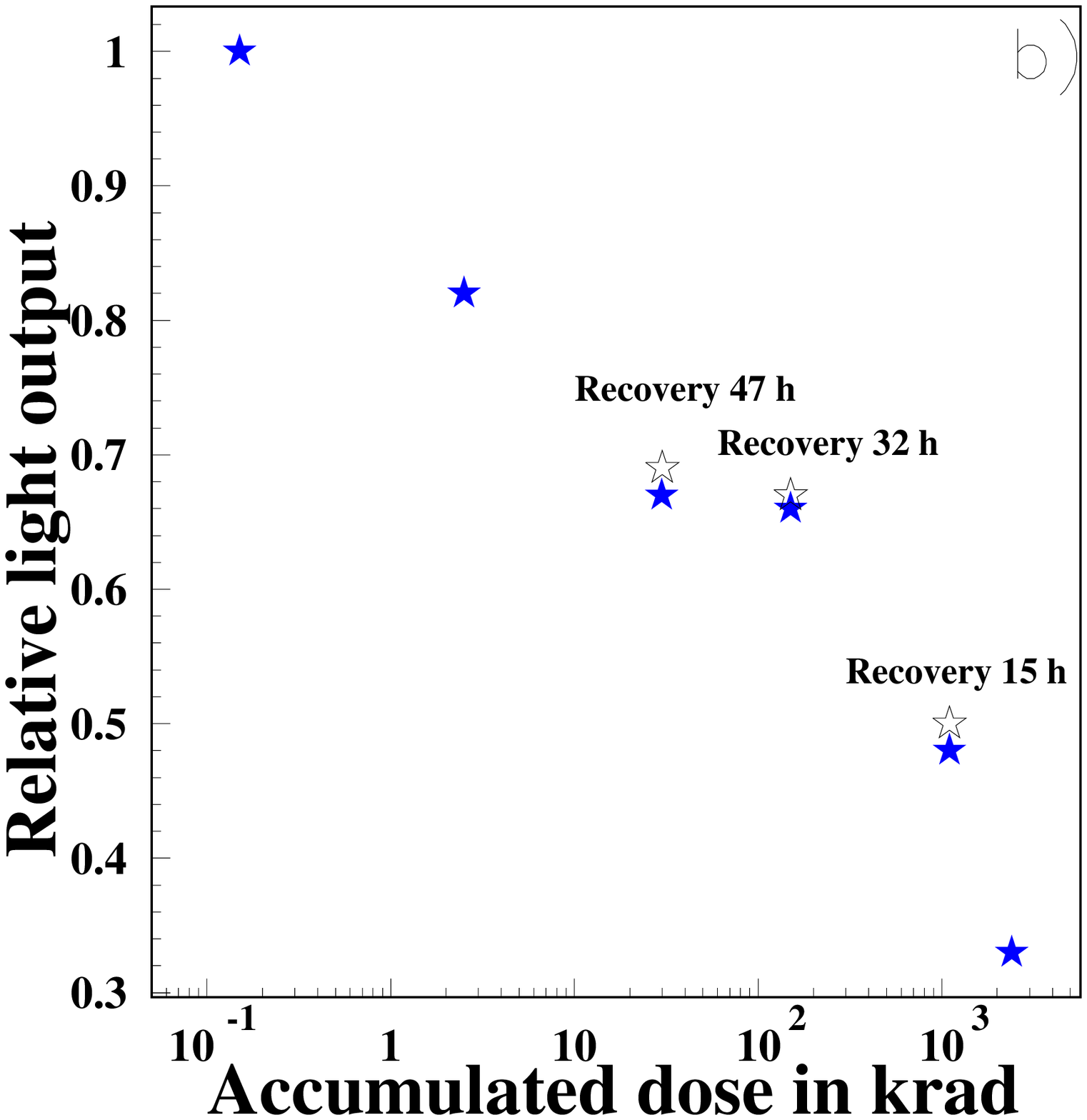,width=85mm,height=70mm}}
\caption{ Light output of (a) Bogoroditsk crystal B21 and (b) Shanghai
crystal S25 in the first exposure versus absorbed 
dose in krad at the 27 GeV electron beam after different steps of irradiation procedure. Low points at each absorbed dose stand for light output just after 
the irradiation. Upper points (if any exist) stands for light output after 
some recovery time. }
\label{fig:bogoroditsk}
\end{figure}

The results of the irradiation of the two crystals in the first exposure 
are presented in Fig.~\ref{fig:bogoroditsk}. 
The procedure was to irradiate the crystals and then measure their light
output immediately thereafter using the 27 GeV electron beam. In some
cases, we measured the light output again after letting the crystals sit
without any radiation. 
The Bogoroditsk crystal (see Fig.~\ref{fig:bogoroditsk}(a) ) lost
$33\%$ of the initial signal after first 3.4~krad dose. After the second irradiation,
the absorbed dose increased up to 43~krad and the signal loss was
increased up to $46\%$.
After 47 hours of recovery time, the signal rose up to $70\%$. After 
the third irradiation, the Bogoroditsk
crystal accumulated 313~krad and the signal was at the level of 49$\%$.
32 hours of recovery time returned it to a level of 57$\%$. After
the fourth dose  the total radiation was 2300~krad and the signal level 
was at 37$\%$.
After 15 hours recovery time, the signal recovered slightly to 39$\%$.\\

The Shanghai crystal (see Fig.~\ref{fig:bogoroditsk}(b) ) lost
$18\%$ of the signal after first 1.8~krad dose. After the second irradiation,
the integrated dose increased up to 23~krad and the signal loss
increased up to $33\%$ relative to the signal before the irradiation.
After 47 hours of recovery time, the signal rised up to $69\%$. After 
the third irradiation, the Shanghai
crystal accumulated 163~krad and the signal was at the level of 66$\%$.
One should mention that the signal was pretty stable between the second and
the third irradiations including recovery time and was at the level 66-69$\%$ 
for the absorbed doses of 23-163~krad.
After getting 2800~krad, the signal dropped down to 33$\%$. \\

\begin{figure}[ht]
\flushleft{\psfig{figure=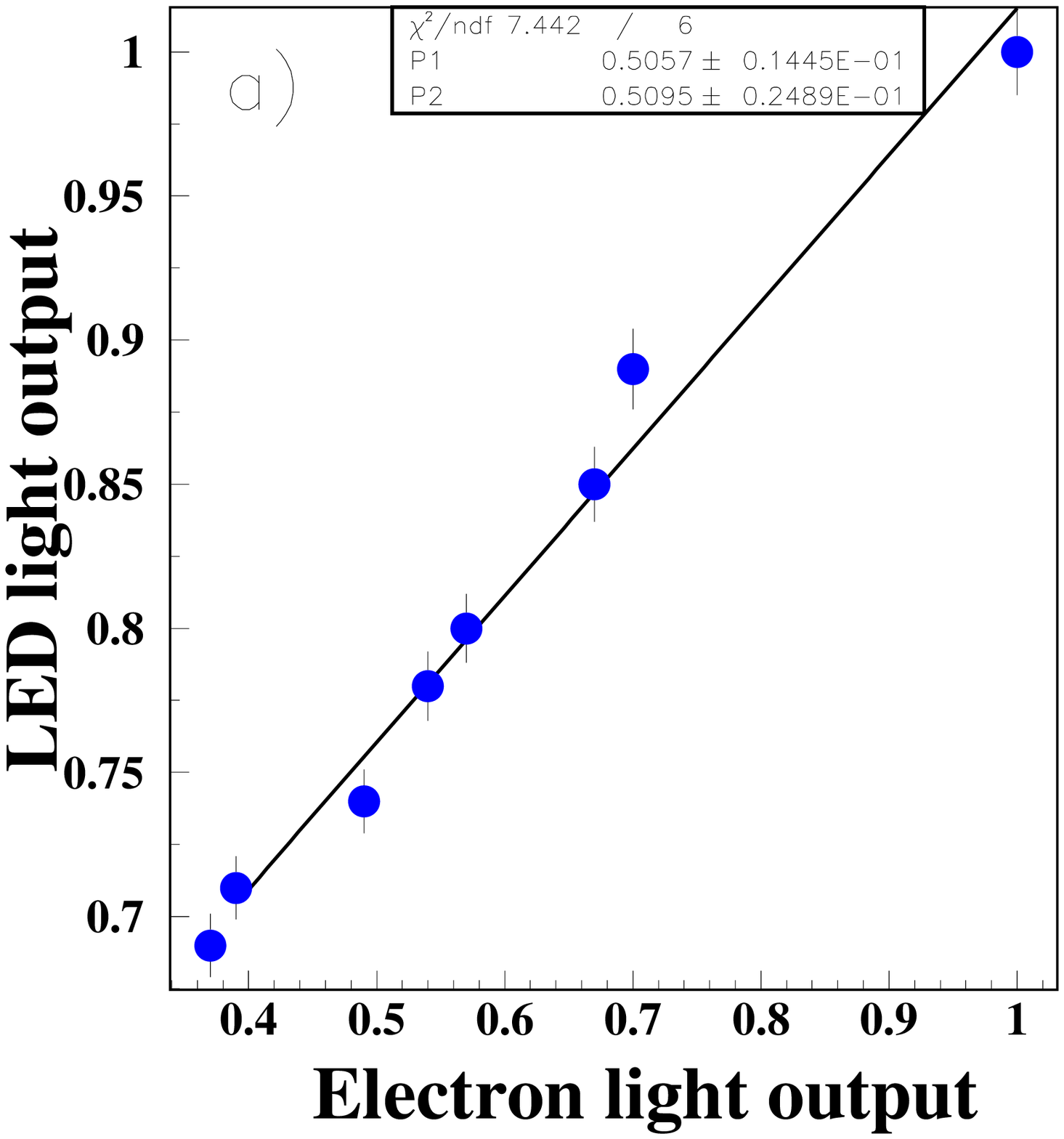,width=85mm,height=70mm}}
\vspace {-73mm}
\flushright{\psfig{figure=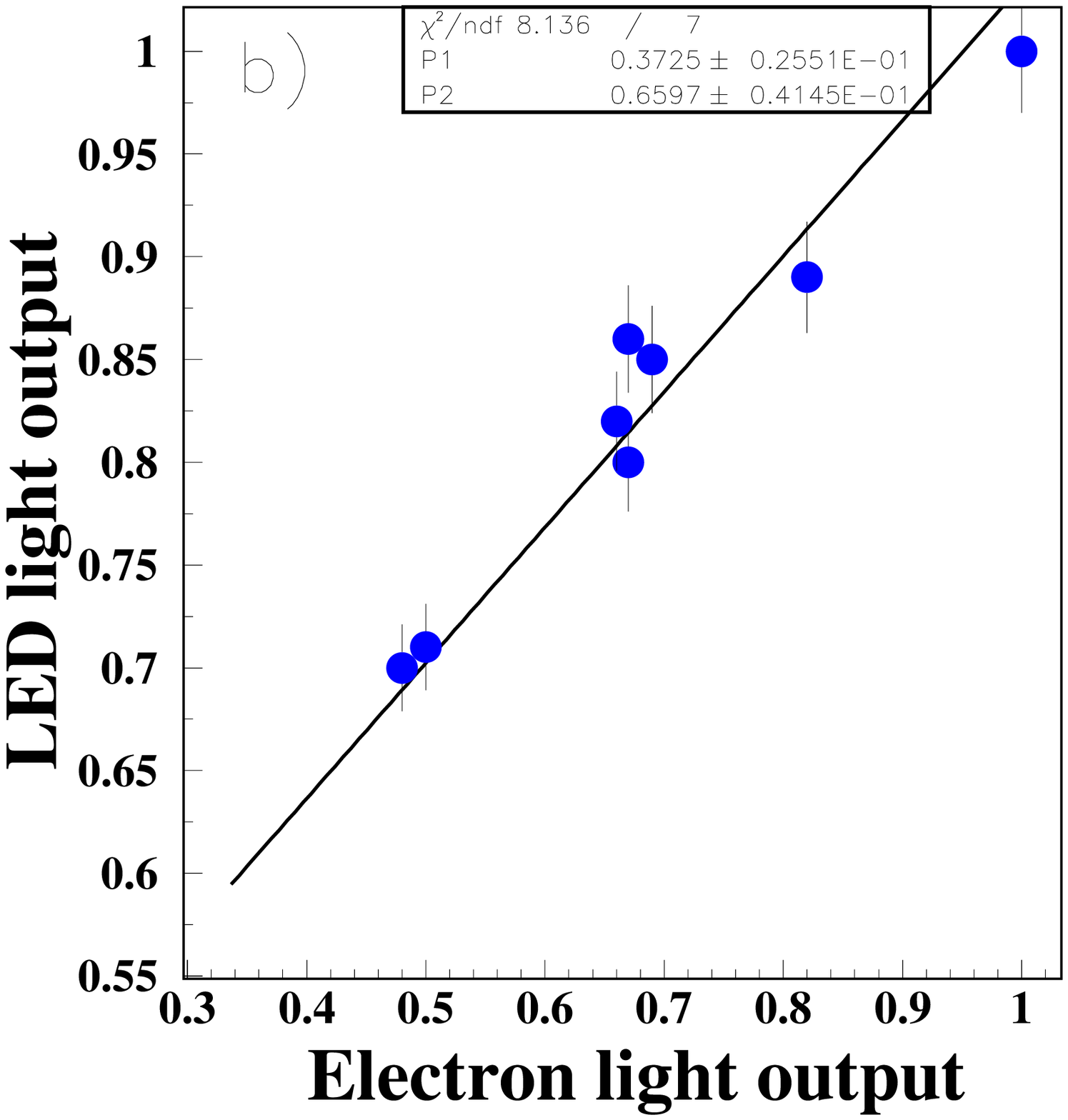,width=85mm,height=70mm}}
\caption{ Correlation between the LED and the beam electron signals for 
(a) Bogoroditsk crystal B21 and (b) Shanghai crystal S25. }
\label{fig:e_led_rus}
\end{figure}

One of the most important conclusions of this work is that even after
an integrated dose about 2.5~Mrad obtained with a super-intensive 
dose rate 100~krad/h both crystals remained usable, although they lost 
2/3 of their light. In BTeV we expect that only $0.1\%$ of the crystals will
receive this much dose in a year.
As was expected, the
constants of proportionality in the MIP-Electron correlations for both the crystals are about 1.
The LED-Electron correlations for both the crystals are shown in 
Fig.~\ref{fig:e_led_rus}. The
constants of proportionality for the two crystals are 0.5 and 0.66.
The degradation of single crystal energy resolution for 27 GeV electrons  
was only $20\%$ for Bogoroditsk crystal and $50\%$ for the Shanghai crystal.\\
 
The four crystals irradiated in the second exposure with a dose rate of 0.5-1 krad/h,
and a total dose of 350-700 rad, lost up to 10$\%$ of their light output for Shanghai
crystals and up to 25$\%$ for Bogoroditsk crystals. After each of the next
three runs no signal loss was seen, within the 3$\%$ accuracy (the systematic
error due to a PMT gain change effect).\\

\section{Summary and Conclusions}

  Radiation hardness of lead tungstate crystals is an important issue
for the BTeV experiment at Fermilab. Simulation of absorbed dose profiles 
in the crystals with the use of the MARS program has shown that 
the dose rates for the crystals range from 0.1 up to 700 rad/h. About 95$\%$ of the crystals in the BTeV electromagnetic calorimeter will 
get the absorbed doses from 0.1 to 30 rad/h assuming that the Tevatron luminosity 
is $2\times 10^{32} cm^{-2}sec^{-1}$. Almost 5$\%$ of the crystals will get from
30 rad/h up to 200 rad/h, and only 0.5$\%$ more than 200 rad/h. \\

  A study of radiation damage in lead tungstate crystals has been carried
out in Protvino in 2001-2002 for BTeV. The crystals were manufactured in Bogoroditsk
(Russia) and Shanghai (China) at the very end of 2000, and in Apatity (Russia)
in early 2002. There were two approaches in the study. First, crystals 
were irradiated  by high-intensity high-energy 
electron and hadron beams 
at radiation doses ranging 
from 0.1 to 60 rad/h. Secondly, crystals were irradiated by charged hadrons,
$\gamma$-quanta and neutrons from the internal target of the U70
in a wide energy spectrum from 10 eV up to 70 GeV 
at dose rates between 0.5 and 100 krad/h.  \\

  The dependence of light output loss on a dose rate has been 
measured. The light loss exhibited saturation when the dose rate was kept
constant. At larger dose rates, the light output loss still saturates but at lower 
light output levels. Each crystal had a different percentage of 
light loss when it saturated. More quantitatively: 
no light output loss was observed for dose rates less than 1 rad/h.
For dose rates of 10-25 rad/h with 27-GeV electron irradiation, eight crystals
lost on the average 8$\%$.
For 40 GeV pions this average was 12$\%$ at comparable irradiation dose rates. 
The difference between the damage due to electron and pion irradiation 
can be attributed entirely to
their difference in the radiation profile along the length of the crystal. Much of electron 
energy is deposited near the shower maximum, from
4 to 10 cm from the front of the crystal. For pion beams, the radiation dose 
profile reaches its maximum around 5-7 cm and stretches all the way to the rear-end 
of the crystal. However, a possible effect due to the difference between 
the physical processes by which electrons and pions interact with 
crystals cannot be ruled out. \\    

For dose rates of 30-60 rad/h using 40-GeV pion irradiation, five crystals
lost on the average  20$\%$.
For a dose rate of 500 rad/h using irradiation by charged hadrons,
$\gamma$-quanta and neutrons with the average energy of 10 GeV, two crystals lost 10$\%$,
and two other crystals lost 25$\%$ when they were exposed to 1 krad/h of radiation.
Two crystals 
got extremely high dose rate of
100 krad/h and accumulated about 2.5 Mrad absorbed dose (maximum annual dose of any BTeV crystals!) also with the same mixed
particle spectra irradiation. They remained useable. Their 
light output loss was a factor of 3. This is far 
from the BTeV environment, where 700 rad/h will be the highest 
$0.1\%$ crystals. \\

There is a correlation between a change in the LED signal and a change
in the beam (electron or MIP) signal under irradiation. The constant of proportionality
is different for different crystals and varies from 0.3 to 0.6 for electron
irradiation and from 0.5 to 0.9 for pion irradiation. 

The non-uniformity (maximum 0.5$\%$ per cm at one third of the crystal
length) of the light yield does not 
change significantly 
when the dose rate is up to 60 rad/h. After 2.5 Mrad absorbed dose with a 
dose rate of 100 krad/h the uniformity became 1.5 times poorer, at least 
for one of our crystals.

When irradiation decreases or stops, crystals recover. 
The average recovery time for six crystals which lost from 7 to 20$\%$ of the LED
signal, is (200$\pm$40) hours, and the damage recovery
after 400 hours was (87$\pm$5)$\%$. \\                          

To summarize, lead tungstate crystals lose light from irradiation 
by high-intensity high-energy beams. This loss level depends on dose
rate.  If dose rate does not change, the light loss saturates.
If the dose rate is reduced, the light output recovers. Crystals have 
to be calibrated continuously during the BTeV experiment. We did not see
a significant difference in radiation hardness of the crystals from the
three manufacturers.\\   

\section{Acknowledgments}

    We would like to thank the IHEP management for providing us
a beam line and accelerator time for our 
testbeam studies. Special thanks are due to Fermilab for 
providing the equipment 
for data acquistion. We would like to thank A.P.Bugorsky, O.A.Grachov, 
I.V.Kotov, 
V.P.Kubarovsky and R.Y.Zhu for useful discussions. This work was partially 
supported by the U.S. National Science Foundation and the Department of Energy.

\end{document}